%% file: SSD-RAID-Reliability-2019-Sep-30.tex
\documentclass[9pt,journal,compsoc]{IEEEtran}
\ifCLASSINFOpdf
\else
\fi
\usepackage[normalem]{ulem}
\usepackage{graphicx}
\usepackage{amsmath}
\usepackage{soul}
\usepackage[linesnumbered,ruled]{algorithm2e}
\usepackage{subfig}
\usepackage{relsize}
\usepackage{multirow}
\usepackage{hhline}
\usepackage{dirtytalk}
\usepackage{xcolor}
\usepackage{soul}
\usepackage{longtable}
\usepackage{tabularx}
\usepackage{afterpage}

\usepackage[hyphens]{url}

\usepackage{mathtools}

\usepackage{ragged2e}
\usepackage{blindtext, graphicx}
\graphicspath{{Pictures/}{Profile/}} 
\usepackage{enumitem} 
\usepackage{hyphenat} 
\usepackage{amsmath, amsthm, amssymb}
\usepackage{dblfloatfix}
\usepackage{soul}
\usepackage{xcolor}
\usepackage{cite}

\usepackage{graphicx}
\usepackage{dirtytalk}
\usepackage{subfig}
\usepackage{graphicx}
\usepackage{color}
\usepackage{makecell}
\usepackage[font=scriptsize]{caption}
\usepackage{dblfloatfix}
\graphicspath{ {Figures/} }
\usepackage{multirow}

\usepackage[numbers, square, comma, sort&compress]{natbib}

\usepackage{tikz}
\usepackage{xcolor}

\definecolor{blue}{rgb}{0.0, 0.0, 1.0}

\SetCommentSty{mycommfont}

\newcommand\HWM{{HWM}}
\newcommand\SWM{{SWM}}

\definecolor{amber}{rgb}{1.0, 0.75, 0.0}

 \definecolor{mytxtcolor}{rgb}{0.2, 0.2, 0.6}

\begin{document}
	%
	\title{Evaluating Reliability of SSD-Based I/O Caches in Enterprise Storage Systems}
	
	
	\author{Saba~Ahmadian,~Farhad~Taheri,~Hossein~Asadi,~\IEEEmembership{Senior Member,~IEEE}
		\IEEEcompsocitemizethanks{\IEEEcompsocthanksitem Saba Ahmadian, Farhad Taheri, and Hossein Asadi are with the Data Storage, Networks, and Processing (DSN) Lab, Department of Computer Engineering, Sharif University of Technology, Tehran, Iran.\protect\\
			E-mail: ahmadian@ce.sharif.edu; farhadtaheri@ce.sharif.edu; asadi@sharif.edu
			
		}
	}

	\IEEEtitleabstractindextext{%
		{\justify
			\begin{abstract}
				I/O caching techniques are widely employed in enterprise storage systems in order to enhance performance of  I/O intensive applications in large-scale data centers.
				Due to higher performance compared to \emph{Hard Disk Drives} (HDDs) and lower price and non-volatility compared to \emph{Dynamic Random-Access Memories} (DRAM), Flash-based \emph{Solid-State Drives} (SSDs) are used as a main media in the caching layer of storage systems.
				Although SSDs are known as non-volatile devices but recent studies have reported large number of data failures due to power outage in SSDs.
				To overcome the reliability implications of SSD-based I/O caching schemes, RAID-1 (mirrored) configuration is commonly used to avoid data loss due to uncommitted write operations.
				Such configuration, however, may still experience data loss in the cache layer due to correlated failures in SSDs.
				To our knowledge, \emph{none} of previous studies have investigated the reliability of SSD-based I/O caching schemes in enterprise storage systems.

				In this paper, we present a comprehensive analysis investigating the reliability of SSD-based I/O caching architectures used in enterprise storage systems under power failure {and high-operating temperature}.
				We explore variety of SSDs from top vendors and investigate the cache reliability in mirrored configuration.
				To this end, we first develop a physical fault injection and failure detection platform and then investigate the impact of workload dependent parameters on the reliability of I/O cache in the presence of two common failure types in data centers, \emph{power outage} and \emph{high temperature} faults.
				{We implement an I/O cache scheme using an open-source I/O cache module in Linux operating system.}
				{The experimental results obtained by conducting more than twenty thousand of physical fault injections on the implemented I/O cache with different write policies reveal that the failure rate of the I/O cache is significantly affected by workload dependent parameters.
				Our results show that unlike workload requests access pattern, the other workload dependent parameters such as request size, \emph{Working Set Size} (WSS), and sequence of the accesses have considerable impact on the I/O cache failure rate.
				We observe a significant growth in the failure rate in the workloads by decreasing the size of the requests (by more than 14X).} 
				{Furthermore, we observe that in addition to writes, the read accesses to the I/O cache are subjected to failure in presence of sudden power outage (the failure mainly occurs during promoting data to the cache).} 
				In addition, we observe that I/O cache experiences \emph{no} data failure upon high temperature faults.
		\end{abstract}}
		
		\begin{IEEEkeywords}
			Flash-Based Solid-State Drives (SSDs), Storage Systems, I/O Cache,  Reliability Analysis, Power Outage.
	\end{IEEEkeywords}}

	\maketitle
	
	\IEEEdisplaynontitleabstractindextext
	
	%
	\IEEEpeerreviewmaketitle

		\section{Introduction}
\begin{figure}[!t]
		\centering
		\hspace*{-.25cm} 
		\includegraphics[scale=0.7]{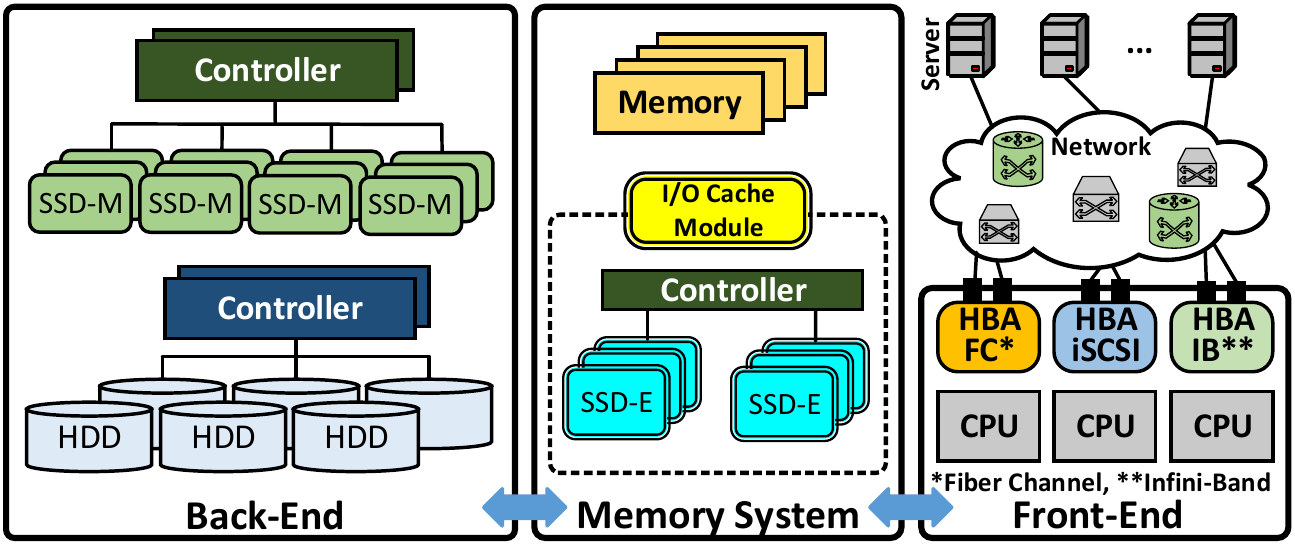}
		\caption{Overview of an enterprise storage system (SSD-M: Midrange-SSD and SSD-E: Enterprise SSD).}
		\vspace{-2em}
		\label{fig:ent-storage-sys}
	\end{figure}

	Increasing the I/O intensive applications such as \emph{Online Transaction Processing} (OLTP) and banking services makes storage subsystems  built upon \emph{Hard Disk Drives} (HDDs) as the performance bottleneck of enterprise systems.
	In order to alleviate the performance shortcomings of HDD-based storage subsystems, enterprise manufacturers such as Dell EMC, NetApp, and HP \cite{emc_fast_cache, netapp_ssd_cache, hp_smart_cache} and emerging storage architectures \cite{ahmadian2018eci,reca,lbica} employ {high performance flash-based devices such as enterprise \emph{Solid-State Drives} (SSDs) as a cache layer for disk subsystem, which is mainly composed from low-performance HDDs and mid-range flash-based SSDs} (as depicted in Fig. \ref{fig:ent-storage-sys}). SSDs are non-volatile devices which because of their non-mechanical design provide higher performance and lower power consumption compared to HDDs \cite{micheloni2012inside,salkhordeh2015operating,reca,li2017workload,7887724}. In addition, SSDs cost about 20X lower than volatile \emph{Dynamic Random-Access Memories} (DRAMs) and also do not require additional peripherals such as backup batteries to retain data in case of power outage \cite{leventhal2008flash}.
	
	Employing SSD-based I/O caches in enterprise storage systems can enhance the performance of I/O intensive applications.
	In such SSD-based I/O cache architecture, however, the SSD cache becomes the single point of failure because of buffering write pending requests where any failure in the SSD device leads to data loss.
	Although SSDs are known as non-volatile devices but recent studies such as \cite{tseng2011understanding, zheng2013understanding,ahmadian-ssd-rel-date} have reported different types of failures such as data, metadata, and device failures in the SSDs under \emph{power outage}.
	To {enhance} the reliability of I/O cache and reduce the probability of data loss, enterprise storage systems such as Dell EMC and HP employ \emph{Redundant Array of Independent Disks} (RAID) \cite{chen1994raid} in the configuration of I/O caches \cite{emc_fast_cache,hp_smart_cache}. In such architecture, multiple SSDs are typically configured as RAID-1\footnote{RAID-1, also known as mirrored configuration, replicates data blocks in two or more paired devices.} in the cache layer of enterprise storage systems (Fig. \ref{fig:ent-storage-sys}).
	Such configuration improves both performance and reliability but cannot completely resolve the reliability issues at the RAID level.
	The mirrored configuration keeps two copies of each data in two different devices known as{ \emph{primary} and \emph{secondary} }disks (i.e., primary is the one that the RAID controller chooses to write first) and provides high level of reliability upon disk failures \cite{chen1994raid,thomasian2005performance,thomasian2006multilevel,thomasian2009higher,thomasian2006mirrored,thomasian2012performance,weddle2007paraid}.
	For each write request coming from the application, two identical write operations are performed in both primary and secondary disks.
	{RAID-1 configuration doubles the read performance of the disk subsystem by involving \emph{only} one of the disks which has a minimum queue size and service time for read operations.}
	Such configuration tolerates \emph{disk failures} in the subsystem  
	while data {failures such as shorn writes (i.e., incomplete writes), flying writes (i.e., misplaced writes), and unserializablity (i.e., out-of-order writes) reported} in \cite{tseng2011understanding, zheng2013understanding,ahmadian-ssd-rel-date,cai2017vulnerabilities} cannot be tolerated completely \cite{chen1994raid,thomasian2005performance,thomasian2006multilevel,thomasian2009higher,thomasian2006mirrored,thomasian2012performance,weddle2007paraid}.
	However, a process namely \emph{inconsistency check} or \emph{scrubbing} runs in the background and regularly checks the consistency of primary and secondary disks in RAID configuration.\footnote{In other types of RAID, consistency check operation checks the correctness of stripes parity.}
	In RAID-1 configuration, in case of inconsistency (i.e., when the comparison of primary and secondary disks fails), the data of primary disk is copied to the secondary disk. Such operation may destroy the correct data {stored} in the secondary disk by {a} faulty data {block} in the primary disk (this issue happens when data failure is occurred in the primary disk) \cite{dell-cc}. 
	
	Recent studies such as \cite{tseng2011understanding, zheng2013understanding,ahmadian-ssd-rel-date} have \emph{only} investigated the impact of power outage on the failures of SSDs while they have neglected the impact of other parameters affecting the reliability of SSDs such as temperature. In addition, recent studies have \emph{only} focused on the reliability of SSDs in non-mirrored (single) configuration under power outage while the {impact of such failures on I/O cache architectures (commonly in RAID-1 configuration)} has been neglected. Furthermore, the previous studies do not {emulate} the real power outage effect that occurs in data centers and ignore the discharge delay of large-size embedded capacitors in \emph{Power Supply Units} (PSUs).
	
	In this paper, we investigate the failures of SSD-based I/O caches in RAID-1 configuration, which is widely used in enterprise storage systems under two common failure types in data centers, i.e.,  \emph{power outage} and \emph{high temperature} faults. 
	To do so, we develop a reliability test platform including Hardware-Software codesign which injects real faults that may occur in data centers such as power outage and high temperature.\footnote{Several recent works present test platforms and evaluate the reliability and lifetime of other emerging technologies such as \emph{Thermally Assisted Switching-Magnetic Random Access Memory} (TAS-MRAM), \emph{Resistive Random-Access Memory} (RRAM) \cite{7499876}, and memristors \cite{7494670}. Our work mainly evaluates the reliability of SSDs.}
	{We integrate EnhanceIO as an open-source I/O cache module with the kernel to implement the I/O cache level and analyze the reliability of committed requests to the subsystem in presence of I/O cache.}
	We classify different types of failures into 1) \emph{False Write Acknowledge} (FWA) {in which the data is not written in the SSD while the acknowledgment is received in the application level, 2) \emph{Full Data Corruption} (FDC) in which entire data is corrupted.}	 3) shorn writes (i.e., incomplete writes) {in which only a portion of data is written in the SSD}, 4) flying writes (i.e., misplaced writes) {where the write operation is performed in a wrong address}, 5) unserializability issue {which is due to the concurrent writes from different threads to an identical address}, and 6) I/O error (i.e., failures due to disk unavailability) in the disk subsystem. Furthermore, the proposed test platform measures the temperature, current, voltage, and power consumption of under test SSDs.
	
	Using the proposed physical fault injection platform, we conduct real experiments on more than 10 enterprise SSDs from different vendors.
	In our experiments, we first study the impact of workload dependent parameters such as 1) workload \emph{Working Set Size} (WSS), 2) request size, 3) request type, 4) access pattern, and 5) sequence of the accesses on the ratio of different failures on the I/O cache (including SSDs in mirrored configurations). Second, based on the ratio of detected failures, we propose a comprehensive classification of  failures that occur in  {I/O caches} due to power outage.
	\footnote{{In this work, we evaluate the impact of workload characteristic on the reliability of I/O caches. Our initial experiments reveal different levels of I/O cache reliability for various cache configurations such as write policy (evaluated in this work), block size, and replacement policy. These observations show the importance of assigning efficient cache configuration for the applications based on their workload characteristics to get the maximum reliability under power outage faults. The investigation of {the impact of} cache configuration parameters on the reliability of I/O caches will be considered as a future work of our study.}}
	Our results show that several workload dependent parameters such as request size and workload WSS have significant impact on the failure rate {(about 14X and 44\%, respectively) while the others such as request access pattern does not considerably affect the failure ratio (only by 2\%). 
	In our {experiments, we} observe no data failure in the I/O cache in presence of high temperature faults.
	In addition, the experiments reveal that both read and write requests fail upon power outage in the I/O cache with different write policies such as 1) \emph{Write Back} (WB), 2) \emph{Write Through} (WT), and 3) \emph{Read Only} (RO).}
	
	The main contributions of this work are as follows.
	\begin{itemize}
		\item To our knowledge, this paper is the first to investigate the impact of power outage and high temperature faults which commonly occur in datacenters on the SSD-based I/O cache architectures in enterprise storage systems.\footnote{{In this work, we target the impact of power outage and high operating temperature on the reliability of I/O caches. Other important parameters such as SSD aging can affect the reliability of data in SSDs, which is partially reported in \cite{meza2015large,cai2015data,cai2017vulnerabilities,cai2017error,cai2017errors,cai2012error,cai2013error}.}}
		
		\item We develop a physical reliability test platform to inject power outage and high temperature faults which can distinguish 1) \emph{False Write Acknowledge} (FAW),
		{2) \emph{Full Data Corruption} (FDC),}
		 3) shorn writes, 4) flying writes, 5) unserializable writes, and 6) I/O error failures. The proposed test platform {as the first proposed physical framework} measures the current, temperature, and power consumption of the SSDs during test. 
		
		\item By conducting a set of extensive workloads, we study the impact of workload dependent parameters such as {1) workload \emph{Working Set Size} (WSS), 2) request size, 3) request type, 4) access pattern, and 5) sequence of the accesses} on the reliability of the SSD-based I/O cache architectures {with different write policies such as 1) \emph{Write Back} (WB), \emph{Write Through} (WT), and \emph{Read Only} (RO).} We observe {a} significant impact of workload dependent parameters such as request size and workload {WSS} on the failure ratio in presence of power outage fault.
		In addition, we observe that high temperature faults have no impact on the failure of SSDs in I/O cache layer of storage systems.
		
		\item We conduct real experiments on more than ten enterprise SSDs from different vendors by injecting more than twenty thousand of physical faults and examine the ratio of data failures under different faults. 
		
	\end{itemize} 
	
	
	The rest of the paper is organized as follows. Section \ref{SEC:REL} discusses related work. 
	In Section \ref{sec:failure_type}, we {characterize} different types of failure that occur on the SSDs.
	In Section \ref{sec:proposed}, we present our proposed test platform. Section \ref{SEC:EXPR} provides {our proposed evaluations and observations}. Finally, Section \ref{SEC:CONC} concludes the paper.
	
	\vspace{-1.1em}
	\section{Related Works} \label{SEC:REL}
	In this section, we first provide previous studies that investigate the reliability of SSDs and flash-based devices. Then we present the shortcomings of previous studies and show the key advantages of our study.
	
	Based on the \emph{source of failures}, the previous studies on reliability of flash-based devices can be investigated in two groups.
	The first group analyzes the failure of the flash-based memories that are mainly due to the internal structure of devices. Failures such as endurance, read disturbance, and write disturbance that are originated from the structure of flash-based devices (i.e., there is no external reason for such failures), are studied in the first group. The second group mainly focuses on the failures in the flash-based devices that are mainly due to external events such as power outage and other environmental reasons.
	
	Previous studies such as \cite{meza2015large, Boboila2010a, narayanan2016ssd, grupp2009characterizing,schroeder2016flash,cai2017vulnerabilities,cai2015data,cai2012error,cai2013error,chang2016disturbance} mainly analyze  the failure of flash-based devices such as read disturbance, write disturbance, and endurance. Such failures are commonly reported in chip and device {levels}.
	{SSD failures in Google and Facebook datacenters during six and four years of operation are studied in} \cite{schroeder2016flash,meza2015large}. Meza et al. observed similar  \say{bathtube curve} in the failure trend of SSDs and an additional phase namely \say{early detection}. Life cycle of SSDs consists of four phases: 1) early detection, 2) early failure, 3) usable lifetime, and 4) {wear-out} phase. {Meza} et al. conclude that SSDs in their {wear-out} phase do not experience a monotonic failure rate. In addition to \cite{meza2015large}, other studies such as \cite{narayanan2016ssd} observed more realistic results of SSD failures in production environment.
	
	The other parameters of flash-based memories such as performance, power consumption, and reliability of devices are measured in \cite {Boboila2010a, grupp2009characterizing}. The results reveal a significant difference between the measured and reported values in the datasheets of the products. The measured parameters are used to present the most reasonable trade-off in the design of storage systems.
	
	The impact of power outage {in} embedded systems is investigated in \cite {kim2007virtual}. {A} software-based test platform is proposed which mainly simulates the effects of power outage on the \emph{Flash Translation Layer} (FTL) of SSDs and file systems {at the} \emph{Operating System} (OS) layer. The proposed test platform is programmed to detect a group of previously defined types of failures that occurs during simulation. Such {a platform} neglects the effects of realistic power failures and only is able to detect a limited number of failures that may occur in simulation.
	
	Recent studies such as \cite{tseng2011understanding, zheng2013understanding} have investigated the reliability of flash cells and SSDs under power outage. Such studies have proposed a test platform that injects realistic power outage faults to the devices.  Tseng et al. in \cite{tseng2011understanding} have proposed an FPGA-based reliability test platform {where} the power of under test flash cells is cut by high-speed transistors with less than $3.7\mu s$ delay. The results of the experiments in \cite{tseng2011understanding} {show} several failures in the flash cells (i.e., in the chip level design) due to power outage. Such study investigates the reliability of flash cells in the chip level design which neglects several recovery mechanisms that is employed in device level designs such as SSDs. 
	
	In upper level designs of flash-based devices, most of chip level failures are masked and would not result in data failure in the application layer. However, such designs suffer from other types of failure that may not occur in chip level and hence, investigating the reliability of the SSDs (i.e., device level design) is required. To this end, later studies such as \cite{zheng2013understanding} have evaluated the reliability of SSDs under power outage. 
	
	Zheng et al. have {proposed} a reliability test framework that {explores} the impact of realistic power outage faults on SSDs. They have analyzed fifteen SSDs from five different vendors and reported several failures for thirteen out of fifteen SSDs \cite{zheng2013understanding}. Similar to previous studies, \cite{zheng2013understanding} neglects the impact of workload dependent parameters on the failure ratio where they {only} examine the reliability of SSDs under a fixed and simple workload. 
	In addition, the power failure is performed by high-speed transistors where the input voltage of under test SSDs is dropped to zero in microsecond-order delay. In such condition, the impact of large size capacitors that are provided in \emph{Power Supply Unit} (PSU) is neglected where the SSDs {under test} do not experience the real power outage process that occurs in the systems. Experimental results {in} \cite{ahmadian-ssd-rel-date} show that the fall delay of PSU output (voltage drop from 5v to 0) takes more than 900ms where the SSDs become unavailable {once} the input voltage drops to 4.5v that takes about 40ms. 
	
	The impact of workload dependent parameters on the failure ratio is {examined} in \cite{ahmadian-ssd-rel-date} by injecting realistic power outage failures. However, {\emph{none}} of previous studies investigated the impact of workload dependent parameters on the reliability of SSD-based I/O caches in the enterprise storage systems in the presence of realistic faults. In addition, the proposed test platform in \cite{ahmadian-ssd-rel-date} does not provide the current, temperature, and power consumption of the under test SSDs. Furthermore, \cite{ahmadian-ssd-rel-date} neglects the impact of other common {types} of faults such as high temperature on the failure of SSDs in datacenters.
	
	
		\section{Failure Characterization in SSDs}
	\label{sec:failure_type}
\begin{figure}[!h]
		\centering
		\includegraphics[scale=0.78]{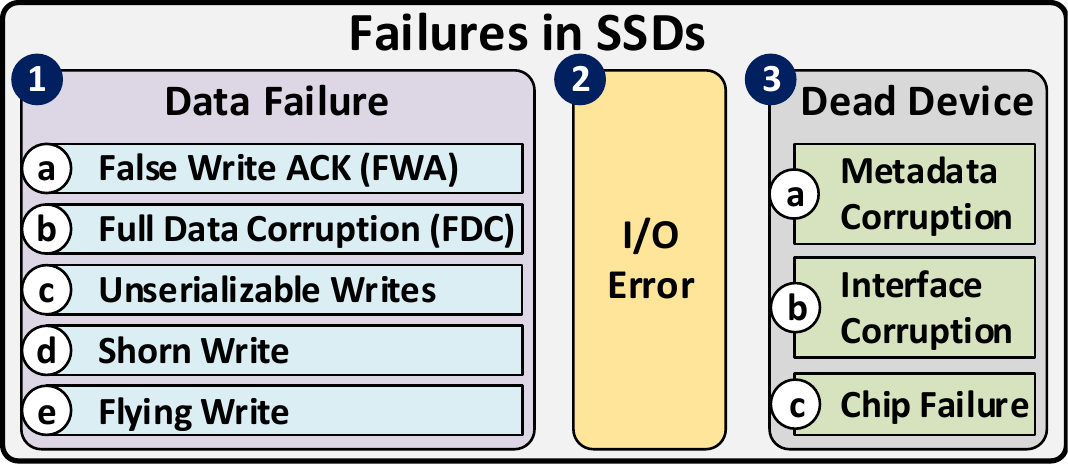}
		 \vspace{-0.5em}
		\caption{Different types of SSD failures.}
		 \vspace{-2em}
		\label{fig:types_of_failure}
	\end{figure}

	{In this section, we elaborate different types of failures that occur in case of either power outage or high temperature (in Section \ref{sec:failure_detection}, we will elaborate on how we detect the failures using our proposed reliability test platform).}
	Fig. \ref{fig:types_of_failure} shows {SSD} failures and {their} relation. As shown in {Fig. \ref{fig:types_of_failure},} the failures are categorized in three groups namely: 1) \emph{data failure}, 2) \emph{I/O error}, and 3) \emph{dead device}. \emph{Data failures} may occur in five types: 1) \emph{False Write Acknowledge} (FAW), 2) \emph{unserilizable writes}, 3) \emph{shorn writes}, 4) \emph{flying writes}{, and 5) \emph{Full Data Corruption} (FDC)}. In the {following,} we describe each type of failure and demonstrate how such failures occur.
%
%
%
%
%
%

	\subsection{Data Failure}
	\label{sec:data_failure}
	{Data failure occurs when the correct data becomes unavailable after a successful write operation on the SSD. In the following, we investigate data failures in five different groups. }
	\begin{enumerate}
		
		\item \textbf{False Write Acknowledge (FWA)}:
		This failure occurs when we receive ACK in {the} application layer but the data is not actually written in the SSD due to power outage.
		{This failure is mainly due to the volatile storage elements in the internal data path within the control layer of SSDs such as {\emph{host FIFO buffer}} and {\emph{DRAM cache}}} \cite{cai2017error,cai2017errors,fifocomparison,meza2015large,luo2018architectural}.
		
		\item \textbf{Unesrilizable Writes}:
		{The most common \emph{host controller interface} which schedules the requests  in SSDs is \emph{Advanced Host Controller Interface} (AHCI)} \cite{ahci_intel,luo2018architectural}. {AHCI keeps I/O requests received from OS in a single command queue namely \emph{Native Command Queue} (NCQ) in which the requests are queued out-of-order to provide higher performance} \cite{luo2018architectural}.
		{Such condition raises challenging issues such as unserializable writes where the random and unknown order of committed writes on the SSD fail in presence of power outage which leads to data loss.}
		
		\item \textbf{Shorn Write}:
		{The SCSI layer of the Linux kernel partitions each request into smaller sub-requests and commits them into the disk subsystem.
			In this case, the sub-requests are kept in {\emph{DRAM buffer}} in the SSD internal data path where sudden power outage may corrupt one or more of the sub-requests.
			In this type of {failure, some} parts of an I/O request are written to the SSD (NAND flashes) while others are not.}
		
		\item \textbf{Full Data Corruption (FDC)}:
		{In this type of failure, although the write operation is completed in the SSD but the data block (including all sub-requests) is corrupted. Such failure differs from shorn writes since all sub-requests are failed. In addition, due to  different initial data and final written data, such failure cannot be considered as FWA. {This} failure can be due to  1) {errors in} volatile storage elements in the SSD internal data path (such as {\emph{host FIFO buffer}} and {\emph{DRAM cache}}} \cite{cai2017error,cai2017errors,fifocomparison,meza2015large,luo2018architectural}) {and 2) {errors} in NAND flashes within the SSD such as program errors} \cite{luo2018architectural}.
		
		\item \textbf{Flying Write}:
		{This failure mainly occurs in HDDs, however, it may also occur in SSDs.
			Flying writes in SSDs may be due to the corruption of the SSD mapping tables which are kept in internal DRAM to provide higher performance} \cite{meza2015large,fifocomparison}.
			{In other words, in such type of failure, the correct data is written successfully on a wrong address in presence of power outage.} 
	\end{enumerate}
	
	\subsection{I/O Error}
	\label{sec:io_error}
	This type of {failure} is due to disk unavailability through application during power outage. In this case, some of requests are committed to the disk and ACK is received while we do not receive ACK for the remaining requests. Such requests are blocked until {the} disk {becomes} available and then committed to the disk {subsystem. In case} of long {unavailability,} the application receives a timeout response from  {the} disk subsystem.
	Data failure is occurred for the first group that the ACK is received but because of disk unavailability during power {outage,} the data is destroyed.  
	\subsection{Dead Device}
	\label{sec:dead_dev}
	This type of failure occurs when the SSD becomes broken after multiple power outages. In case of dead device failure, the S.M.A.R.T report of the SSD provides details of the problem. Dead device failure can be experienced in three modes namely: 1) metadata corruption, 2) interface corruption, and 3) chip failure. In the {following,} we elaborate different types of dead device failure.
	\begin{enumerate}
		\item \textbf{Metadata Corruption}:
		This type of failure is due to the problems that occurs for FTL of the SSDs after multiple power outage. In such {failure,} the address map and mapping algorithms which {are} done by the FTL {are} disrupted. When this failure {occurs,} some address areas of the SSD become unavailable through application.
		\item \textbf{Interface Corruption}:
		Multiple power outages affect the SCSI interface of SSDs (typically SATA or SAS) and disrupt the operation of this part. When this failure {occurs,} the SSD becomes unavailable  through \emph{scsi scan} commands from the OS. In some cases, such interface problem affects the controller of the motherboard where the other healthy disks become unavailable. Such problem is resolved by system reboot or disconnecting the faulty SSD from the system. In this type of failure, the S.M.A.R.T report of the SSD includes {\say{\emph{scsi error badly formed scsi parameters}}} log.
		\item \textbf{Chip Failure}:
		In this type of {failure,} the flash chips or internal connections of the SSD are disrupted where the SSD is not recognized by any controller.
	\end{enumerate}
	
	\vspace{-1.2em}
	\section{Proposed Reliability Test Platform} \label{sec:proposed}
	
\begin{figure}[!h]
		\centering
		\includegraphics[scale=0.85]{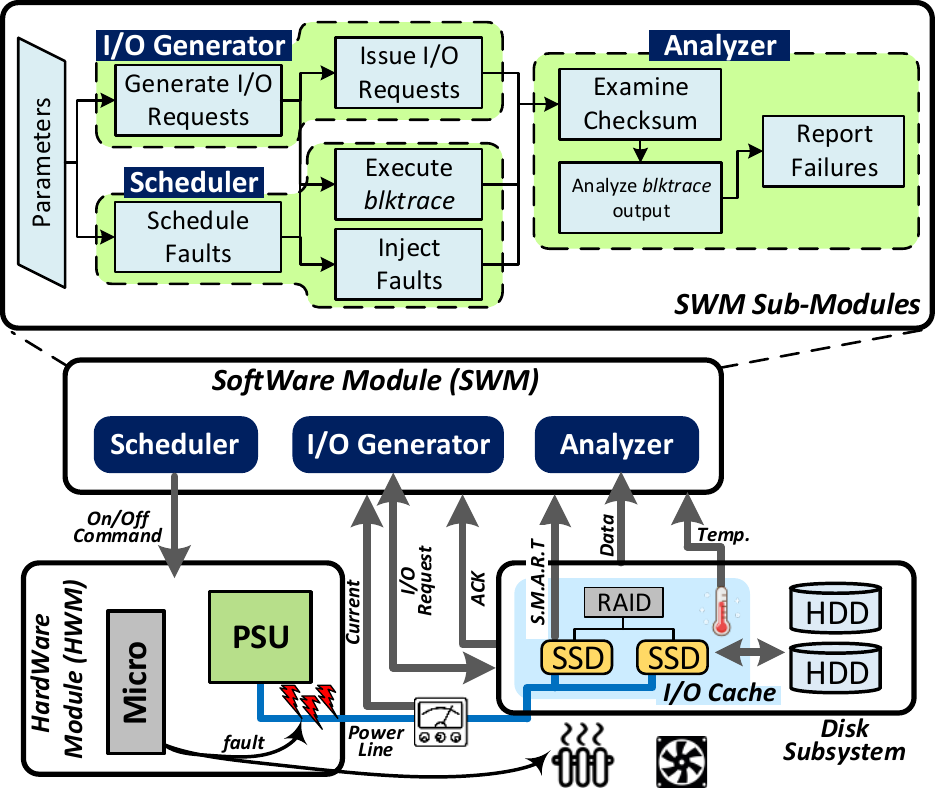}
		\caption{Overview of the proposed test platform.}
		\label{fig:overview}
	\end{figure}

	The {proposed} reliability test platform consists of \emph{HardWare Module} (\HWM{}) and \emph{SoftWare Module} (\SWM{}) working together where \HWM{} is programmed and controlled by \SWM{}. Fig. \ref{fig:overview} shows an overview of  {our} proposed reliability test platform. 
	\SWM{} is responsible for generating the I/O requests and scheduling fault injections time intervals. \HWM{} is programmed to receive commands from \SWM{} and {inject} physical faults to the SSDs. Finally, \SWM{} is used to detect data failure and device failures that are occurred due to injected faults.
	In the following, {we first describe \SWM{} and its components in Section \ref{sec:sw} and then
	in Section \ref{sec:failure_detection} we elaborate the proposed failure detection algorithm. Finally, we provide the structure of \HWM{} in our {proposed} reliability test platform in Section \ref{sec:hw}.}
	
	\subsection{SoftWare Module (SWM)}
	\label{sec:sw}
	\SWM{} generates various workloads with defined parameters such as request size, request type, access pattern, sequence of accesses, and WSS and issues the I/O requests to the disk subsystem {(i.e., HDD equipped with SSD-based I/O cache configuration).}\footnote{{SSDs can be configured in different RAID levels, however, RAID-1 (mirrored) is the common RAID configuration for SSD-based I/O caches} \cite{emc_fast_cache,hp_smart_cache}.} \SWM{} tracks the issued I/O requests and detects the parameters such as issue time, completion time, request size, request type, and the checksum of written data on disk. In addition, \SWM{} keeps the information of the I/O requests in a database in two conditions: 1) before issuing the request to the disk subsystem and 2) after issuing the I/O to the disk subsystem.
	\footnote{{Our proposed reliability test platform also is able to execute real storage workloads, in which the information about requests including 1) size, 2) address, 3) type, and 4) issue time are extracted from workload trace.}}
	\SWM{} is designed to detect different types of failures (as discussed in Section \ref{sec:failure_type}) based on the information that keeps in the database and the header of the I/O requests. \SWM{} {1) schedules fault injection intervals, 2) manages \HWM{} to inject physical faults, and 3) receives the information about SSDs power consumption, temperature, and current.}
	
	{As shown in Fig. \ref{fig:overview}, {three main sub-modules in \SWM{}} namely a) \emph{Scheduler}, b) \emph{I/O Generator}, and c) \emph{Analyzer} perform 1) scheduling faults, 2) submitting I/O requests to the disk subsystem, and 3) detecting failures{, respectively}.} In the following, we describe how each part of \SWM{} works in more details.
	
	\subsubsection{\emph{Scheduler}}
	\emph{Scheduler} is responsible for determining the fault injection intervals. It chooses random time instances that fault injection will be occurred. \emph{Scheduler} communicates with \HWM{} and sends \emph{On/Off Commands} to \HWM{}. It receives information such as 1) temperature, 2) current, and 3) power consumption of the SSDs from \HWM{}.  \HWM{} waits to receive the commands from \SWM{} and turns the SSDs {\emph{on} or \emph{off}} based on \emph{Scheduler}. In addition, the \emph{Scheduler} sets the temperature of the under test SSDs where \HWM{} increases or decreases the temperature of the SSDs. Based on \emph{Scheduler's} command, \HWM{} turns on or off the heater or {the} fan to regulate the temperature of the test platform.
	
	\subsubsection{\emph{I/O Generator}}
\begin{figure}[!h]
		\centering
		\includegraphics[scale=0.48]{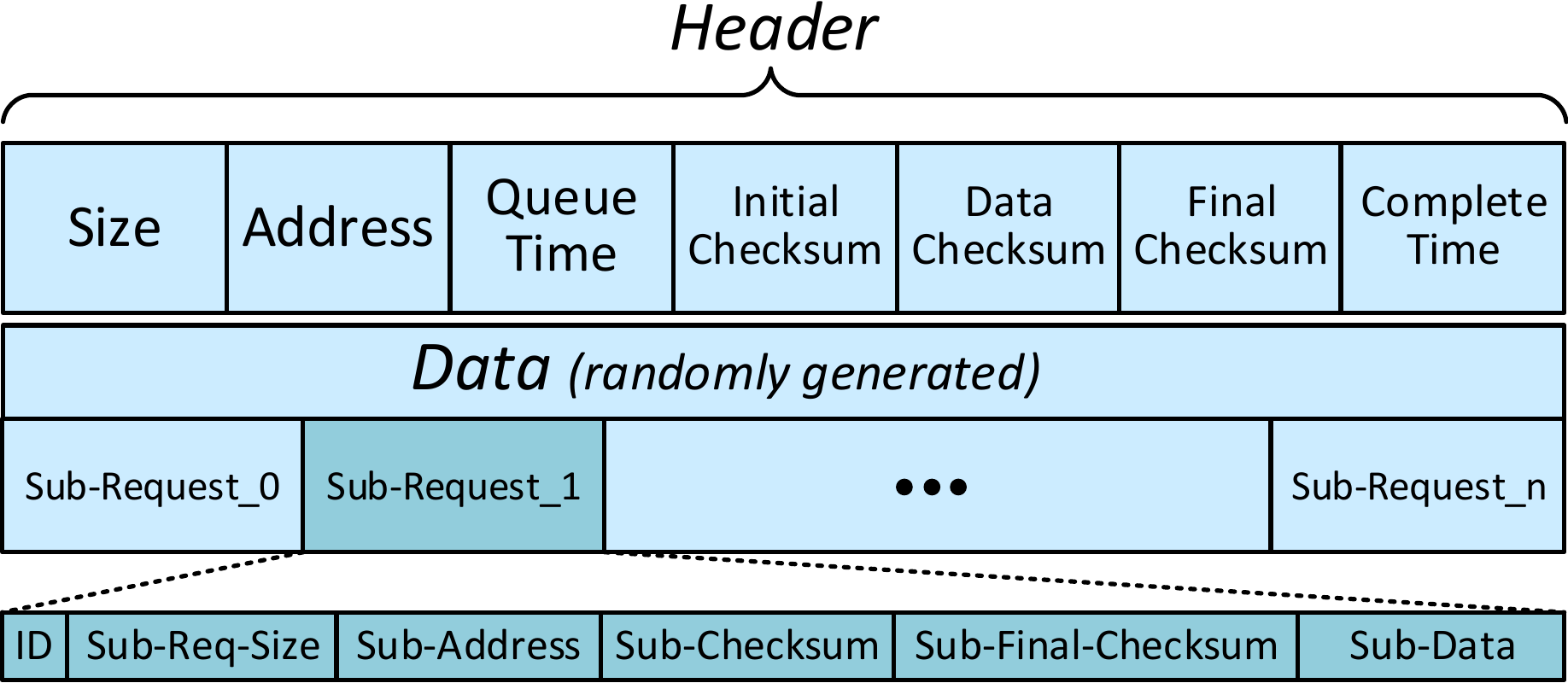}
		
		\caption{Structure of \emph{data packets}.}
		
		\label{fig:data_packet}
	\end{figure}

	\emph{I/O Generator} creates different {types} of workloads with different parameters such as 1) request size, 2) request access pattern, 3) request type, 4) \emph{Working Set Size} (WSS), and 5) sequence of the accesses{. Next, it} commits the I/O requests of the workloads to the disk subsystem. 
	The generated requests are named as \emph{data packets} that are issued to the SSDs by the \emph{I/O Generator}. The structure of \emph{data packets} is shown in Fig. \ref{fig:data_packet}. It can be seen that {a} \emph{data packet} {consists} of two parts: 1) header and 2) data. The data which can include \emph{sub-requests} is produced randomly and the header includes the key information about data such as: 1) size, 2) destination address, 3) issue time (i.e., when the request {enters into the} disk queue), 4) completion time (i.e., when the ACK of the request is received in the application layer), 5) checksum of data before issuing the request, and 6) checksum of written data to the disk subsystem {(Table \ref{detail_packet} provides the detailed information about the content of data packet)}. The information in the header of \emph{data packet} is used in detecting different types of failures (in future we elaborate the failure detection mechanism). Each \emph{sub-request} included in  {the}  data part of \emph{data packet} consists of 1) \emph{ID}, 2) \emph{size}, 3) \emph{address}, 4) \emph{checksum}, and 5) \emph{data} (as depicted in Fig. \ref{fig:data_packet}). 
	\emph{Data packets} are stored in a database to be used in failure detection phase.
	
	\begin{table}[!h]
		\centering
		\caption{Description of fields in data packet.}
		\label{detail_packet}
		\scriptsize
		\begin{tabular}{|l|l|}
			\hline
			\textbf{Field}            & \textbf{Description}                                                                   \\ \hline\hline
			Size             & The total size of I/O request.                                                \\ \hline
			Address          & Destination address of the I/O request.                                       \\ \hline
			Queue\_Time       & When the I/O request enters in disk queue.                                      \\ \hline
			Initial\_Checksum & {\begin{tabular}[l]{@{}l@{}}The checksum of destination address before issuing\\ the request.\end{tabular}}
			
			           \\ \hline
			Data\_Checksum    & The checksum of I/O request (including Data part).                            \\ \hline
			Final\_Checksum   & {\begin{tabular}[l]{@{}l@{}}The checksum of destination address after receiving\\ ACK\end{tabular}}

			\\ \hline
			Complete\_Time    & When the application receives ACK from disk.                       \\ \hline
			Sub\_Request\_i  & The i-th part of the request.                                                  \\ \hline
			ID               & The sub-request number.                                                    \\ \hline
			Sub\_Req\_Size   & The size of sub-request.                                                      \\ \hline
			Sub\_Address     & The destination address of sub-request.                                       \\ \hline
			Sub\_Checksum    & The checksum of sub-request (only sub-data).            \\ \hline
			Sub\_Final\_Checksum    & The checksum of sub-request after completion.            \\ \hline
			Sub\_Data        & The data part of sub-request.                                                 \\ \hline
		\end{tabular}
	\end{table}
	
	\subsubsection{\emph{Analyzer}}
	\emph{Analyzer} keeps {the}  track of  I/O requests and verifies the correctness of written data in the disk subsystem. Based on collected I/O traces, \emph{Analyzer} compares the checksum of \emph{\say{completed}} requests with the stored checksum of corresponding \emph{data packet} in the database. In case of any inconsistency, the \emph{Analyzer} reports a \emph{data failure}.
	\emph{Analyzer} employs \emph{blktrace} as a block layer I/O tracing tool to keep the track of the requests. \emph{Blktrace} is available in Linux kernel (version 2.6 and upper) and provides the information of I/O requests in user level without any performance overhead.
	\emph{Analyzer} detects the other {types} of failures such as device failure and I/O error beside data failures based on the collected information by \HWM{}. In Section \ref{sec:failure_detection}{, we} elaborate how we trace the I/O requests and detect different {types} of failures.
	
	\subsection{Failure Detection}
	
	\label{sec:failure_detection}
	
	In this section, we show how our proposed failure detection algorithm of the test platform detects different types of failure. To do so, we have employed an I/O tracing mechanism which completely traces the I/O requests during running workloads. Such mechanism works online and keeps the state of the I/O requests in different levels. In order to track the I/O requests, we employ the Linux comprehensive I/O tracing tools, namely \emph{blktrace} and \emph{blkparse} providing required details of the request without any performance overhead. In the post-process level, we have employed a modified version of \emph{btt} tool to extract additional information such as standard format of \emph{timing information} of the I/O requests. Such modification helps us in detecting \emph{complete} and \emph{incomplete} I/O requests (we call a request as \emph{complete} when we receive the ACK of the request). To do so, we have modified the operation of \say{--per-io-dump} option in \emph{btt}. This option extracts the trace of {an} individual I/O {request} where such modification creates the trace of large size request that are divided into \say{sub-requests} {with the detailed timing information}. 
	
	Failure detection process works based on the collected information of the I/O requests in the header of the \emph{data packets}. Note that we have two versions of \emph{data packet}: 1) the generated one by the \emph{I/O Generator} kept in database and 2) the written one in the disk subsystem. We start failure detection process in two cases: 1) when we receive a request time out response and 2) when we receive the ACK of a request (i.e., when the \emph{complete} flag of the request is set. A request is set to \emph{complete} when {all its} sub-requests are set as \emph{complete}).
	
	
	{Algorithm \ref{alg:failure-detection} shows how we detect different types of data failures in the platform.
		\footnote{{Detection of dead device failure (mentioned in Section \ref{sec:dead_dev}) is mainly performed in the hardware layer and needs further tests in separated boards and is not performed within SWM.}}
	First, in line \ref{lst:line:check_time_out} we check the timeout response of the request. In case of timeout, we report \emph{I/O Error} in line \ref{lst:line:io-error}. Then in line \ref{lst:line:check_cheksum}, we check the checksum of written data (by comparing $dataChecksum$ and $finalCehcksum$) to detect if data failure is occurred or not. 
	In case of equality, we mark the data as \emph{correct} and report \say{no failure}.
	Otherwise, in case of any inconsistency between $dataChecksum$ and $finalCehcksum$, in line \ref{lst:line:final_not_initial}, we compare the $finalChecksum$ with $initialChecksum$ of corresponding address (which was extracted before write operation and kept in the database) to validate the write operation. In case of equality, in line \ref{lst:checkWAW}, we check whether multiple writes are submitted to the disk subsystem. To do so, we check if the sequence of the accesses is \emph{Write After Write} (WAW) or not. If so, the failure is due to \emph{Unserializability} of two parallel write accesses (line \ref{lst:line:unserial}).
	Otherwise in line \ref{lst:line:FWA} we report \emph{False Write Acknowledge} (FWA) which is due to the fact that data is not written in the SSD flash cells while we receive acknowledgment in application level.
	Then in line \ref{lst:line:check_sub_checksum} we check the correctness of sub-requests and keep the number of failed sub-requests in variable $numOfFailedSubReqs$. In  line \ref{lst:line:check_shorn}, if we find failed sub-requests, we report \emph{Shorn Write}. In line \ref{lst:line:check_all_data} in case of failure of all sub-requests, entire data is corrupted and we report \emph{Full Data Corruption} (FDC).
	We report \emph{Flying write} in line \ref{lst:line:fly} in case of inconsistency of sub-addresses and the address of I/O request. To do so, we scan entire addresses of disk subsystem which takes long time.}

	\begin{algorithm}
		\scriptsize
		\DontPrintSemicolon
		\caption{{Data failure detection algorithm}}
		\label{alg:failure-detection}
		\SetKwProg{Fn}{Function}{ is}{end}
		\KwIn{initialChecksum , dataChecksum, finalChecksum, subChecksum, numOfSubReqs, dstAddress}
		\tcc{timeOut is the default parameter defined in linux kernel for I/O requests.}
		\Fn{failureDetection}
		{
			\tcc{First we check if the timout is occured or not. In case of timeout, IO Error is reported.}
			\If{$responseTime > timeOut$}
			{\label{lst:line:check_time_out}
				Report IO Error.\label{lst:line:io-error}} 
			\tcc{We check the correctness of checksum:}
			\If{$finalChecksum~!=~dataChecksum$} 
			{\label{lst:line:check_cheksum}
				\tcc{In case of inequality, data failure is detected. Then we find the type of data failure.}
				\tcc{We check if data is written in SSD or not?}
				\If{$initialChecksum~==~finalChecksum$} 
				{\label{lst:line:final_not_initial}
					\tcc{In this case, we recieve the ACK in application level while the data is not written in the SSD.}
					\tcc{If we have multiple writes from different threads:}
					\If{$WAW$} 
					{\label{lst:checkWAW}
						Report Unserializibility. \label{lst:line:unserial}}
					\Else{Report FWA.\label{lst:line:FWA}}
				}
				\tcc{We check the correctness of subrequests. $numOfFailedSubReqs$ shows the number of failed subrequests.}
				$numOfFailedSubReqs~=~0$ \label{lst:line:failed_sub_variable}\\ 
				\For{$i = 0 \ to \ numOfSubReqs$} {
					\If{$subFinalChecksum~!=~ subDataChecksum$} 
					{\label{lst:line:check_sub_checksum}
						$numOfFailedSubReqs++$}
				}
				\tcc{If we find any failed subrequest}
				\If{$numOfFailedSubReqs~>~ 0$}
				{\label{lst:line:check_shorn}
					Report Shorn Write.}
				\tcc{If all sub-requests are corrupted we have FDC.}
				\uElseIf{$numOfFailedSubReqs~==~numOfSubReqs$}
				{
					\label{lst:line:check_all_data}Report FDC.}
			}
			\If{$numOfFailedSubReqs~==~0$ and $subAddress$ does not fit in $dstAddress$}
			{\label{lst:line:fly}
				Report Flying Write.}	
		}
	\end{algorithm}

	\subsection{HardWare Module (\HWM{})}
	
	\label{sec:hw}
	\begin{figure*}[!htb]
		\centering
		\subfloat[Schematic of \HWM{} for injecting power outage faults.]{\includegraphics[width=.45\textwidth]{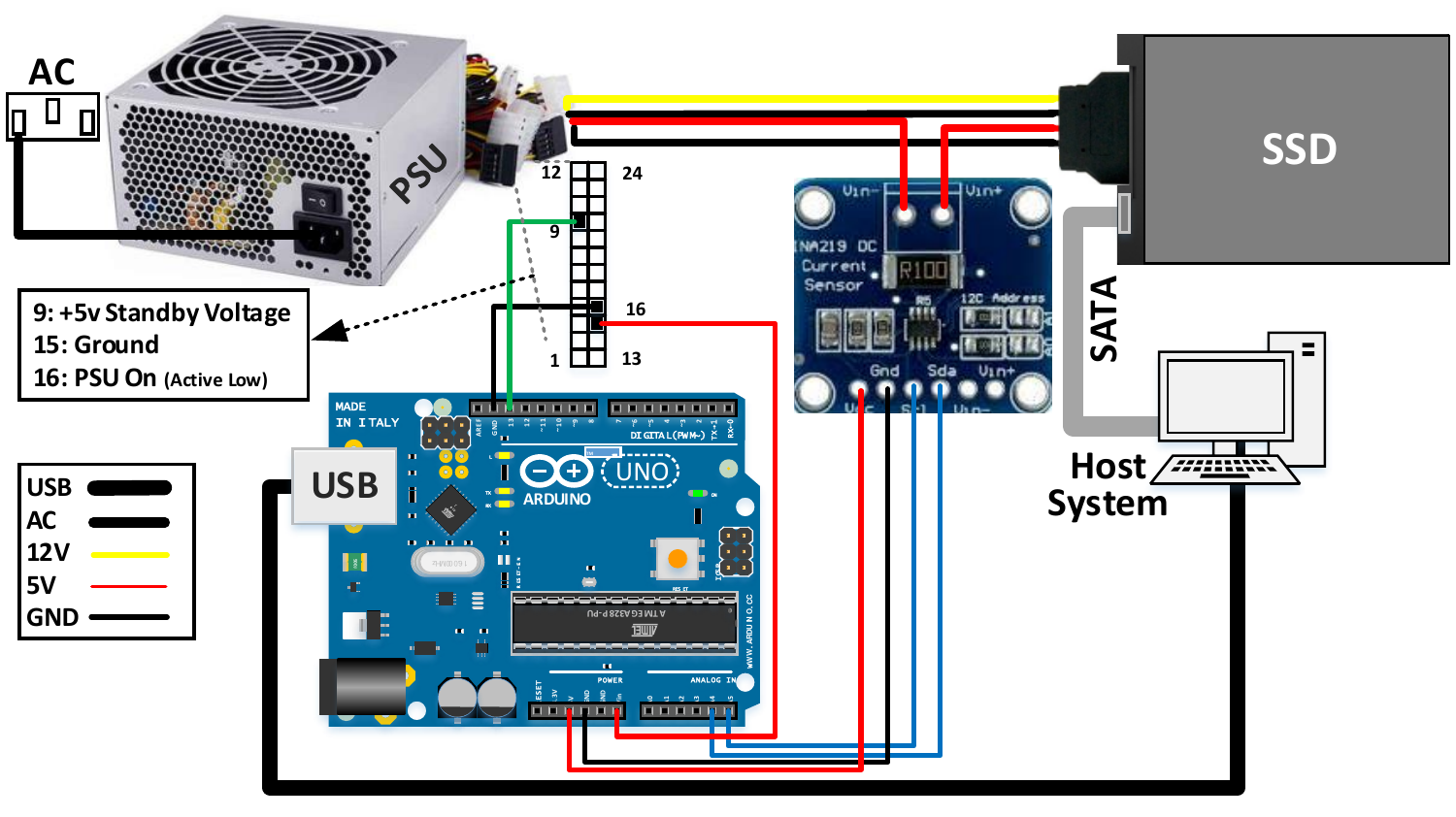}%
			\label{fig:hw_schem}}
		\hfil
		\hspace{-.8pt}
		\subfloat[Schematic of \HWM{} for injecting high temperature faults.]{\includegraphics[width=.45\textwidth]{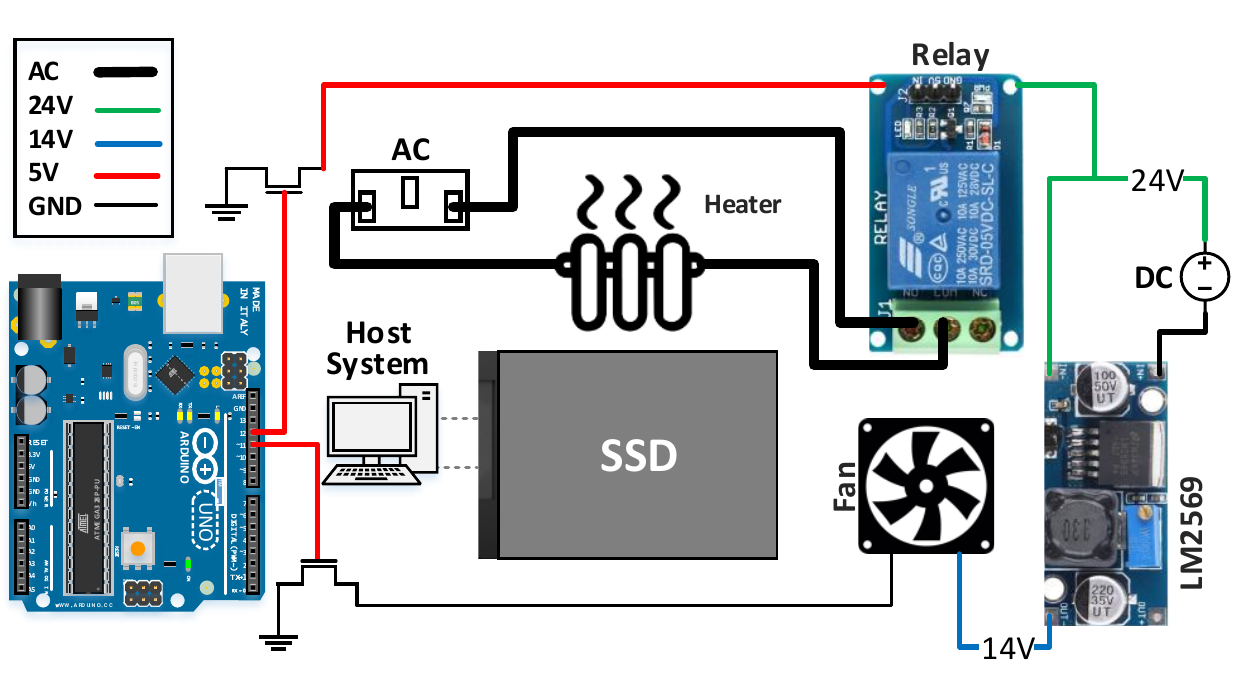}%
			\label{fig:hw1}}
		\hfil
		\hspace{-.8pt}
		\subfloat[Proposed reliability test platform (close view).]{\includegraphics[width=.38\textwidth]{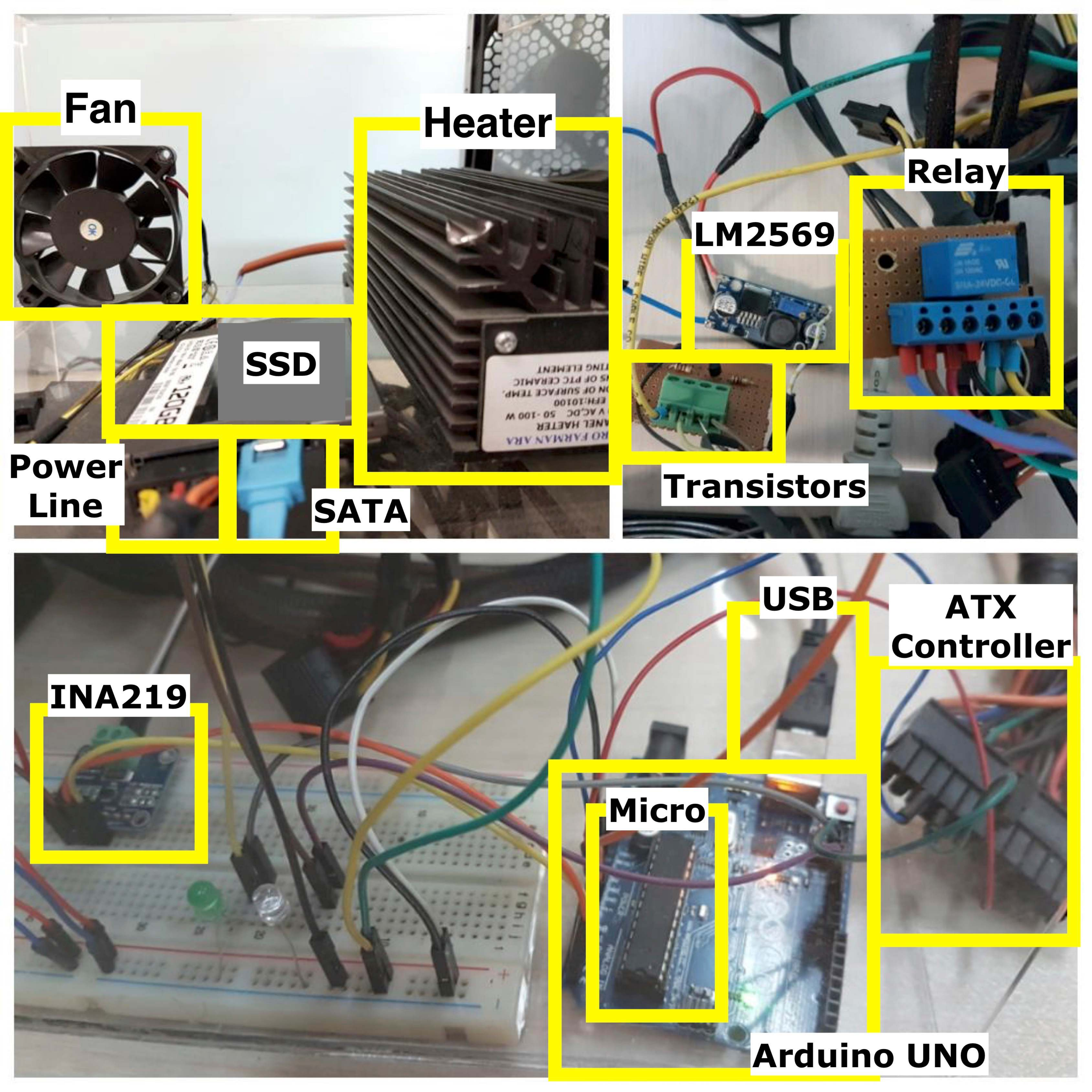}%
			\label{fig:hw2}}
		\hfil
		\hspace{-.8pt}
		\subfloat[Proposed reliability test platform (overall).]{\includegraphics[width=.5\textwidth]{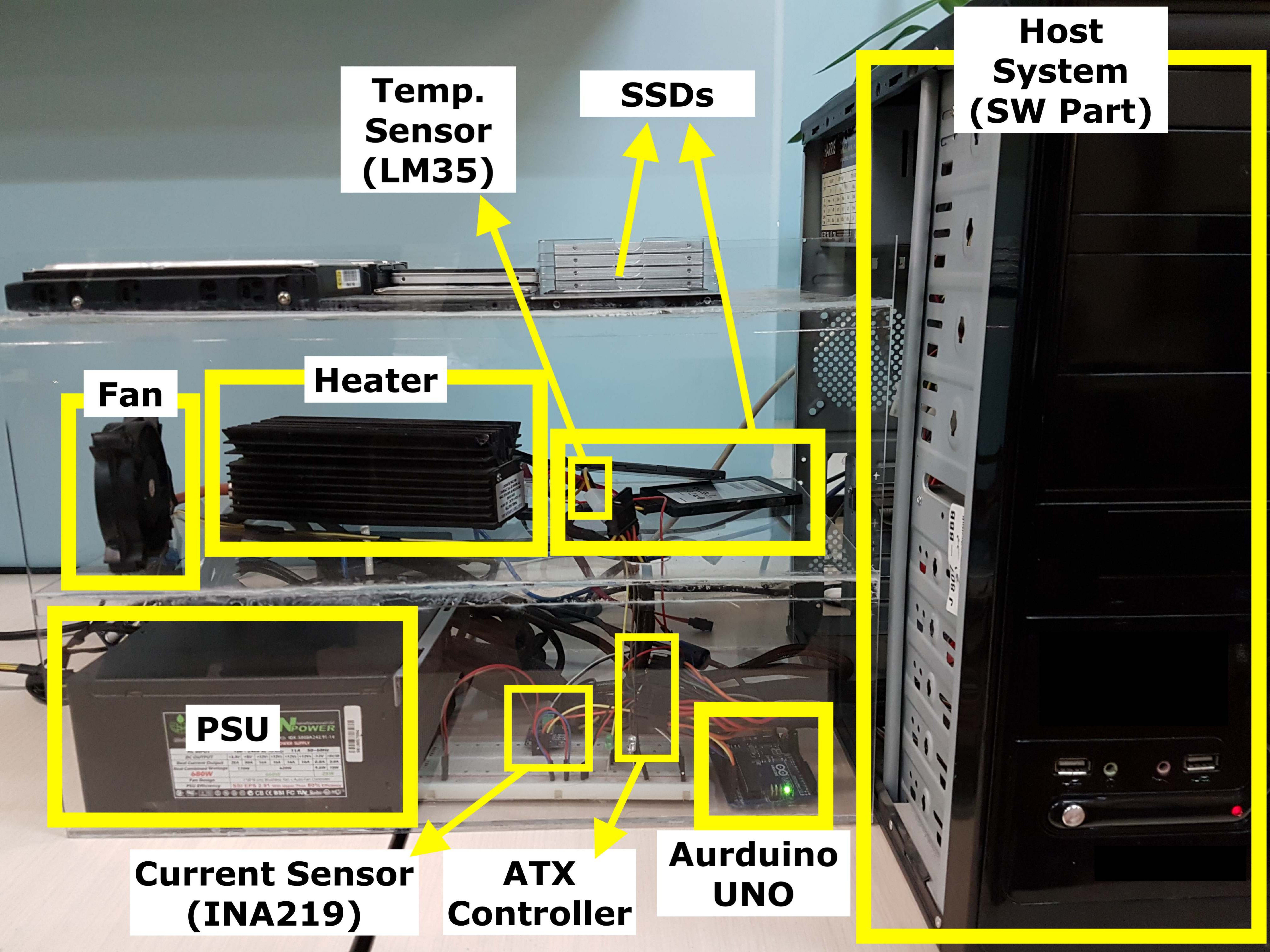}%
			\label{fig:hw-1}}
		
		\caption{Proposed reliability test platform (\HWM{}).}
		
		\label{fig:hardware}
	\end{figure*}
	
	The structure of \HWM{} in our proposed reliability test platform is shown in Fig. \ref{fig:hardware}. \HWM{} is responsible for injecting real and physical faults such as power outage and high temperature failures. \HWM{} receives the fault injection commands from \SWM{} through a USB connection. Fig. \ref{fig:hw_schem} shows the schematic of \HWM{} that injects physical power outage faults to the SSDs. It can be seen that \HWM{} is placed in the path of power lines of the SSDs to perform real fault injections. Fig. \ref{fig:hw1} shows the schematic of  \HWM{} which is responsible for high temperature fault injection. It can be seen that \HWM{} manages the \emph{heater} and {the} \emph{fan} to control the temperature of the SSDs as assigned by \SWM{} (\emph{Scheduler}). The {physical view} of the proposed reliability test platform is provided in Fig. \ref{fig:hw2} and Fig. \ref{fig:hw-1}.
	
	\HWM{} includes the following parts:
	\begin{enumerate}
		\item Atmega32 micro-controller embedded in an Arduino UNO board. This micro-controller {receives and decodes the commands from  \SWM{}.} It turns the power of SSDs \emph{on} or \emph{off} based on the received command in determined time instances. In addition, the micro-controller is programmed to decode the commands related to temperature, fan, and measuring modules.
		
		\item INA219 module is used for measuring the current and voltage of SSDs. This module measures and sends the current of the SSDs to \SWM{} and makes the \SWM{} able to detect the status and operation of the SSDs.
		
		\item SRD24v relay which controls the power of  {the} heater module in the platform. This relay is controlled by the micro-controller and turns the heater \emph{on} or \emph{off} in determined time instances which is assigned by the micro-controller based on the temperature of the SSDs.
		
		\item LM35 sensor measures the temperature of the platform and the SSDs. The information about temperature of the SSDs are passed to \SWM{}.  \SWM{} also {receives} the temperature of the SSDs from S.M.A.R.T report provided by the SSD manufactures. 
		
		\item  {The} heater is responsible for injecting high temperature faults to the SSDs. The employed heater is able to increase the temperature up to 100 degrees Celsius in 5 minutes (in such condition the temperature of the SSDs increases up to 70 degrees Celsius).  
		
		\item  {The} fan is used to cool the platform and decrease the temperature of the SSDs as determined by micro-controller during high temperature {fault} injection.
		
	\end{enumerate}

	\HWM{} communicates with \SWM{} (\emph{Host System}) through a USB serial connection. 
	The embedded micro-controller in \emph{Arduino UNO} receives commands from  \SWM{}. The micro-controller switches the power of SSD to ON or OFF state by controlling the pin  16 of the ATX controller of the PSU which drives the under test SSDs power. Pin 16 of the ATX controller is active low and cuts off the output power of the PSU by applying a high voltage (+5V).
	To inject the high temperature faults, micro-controller manages the heater and fan to control the temperature of the SSDs as \SWM{} decides.
	
\begin{figure}[!h]
		\centering
		\subfloat[When the PSU does not drive any device.]{\includegraphics[width=.45\textwidth]{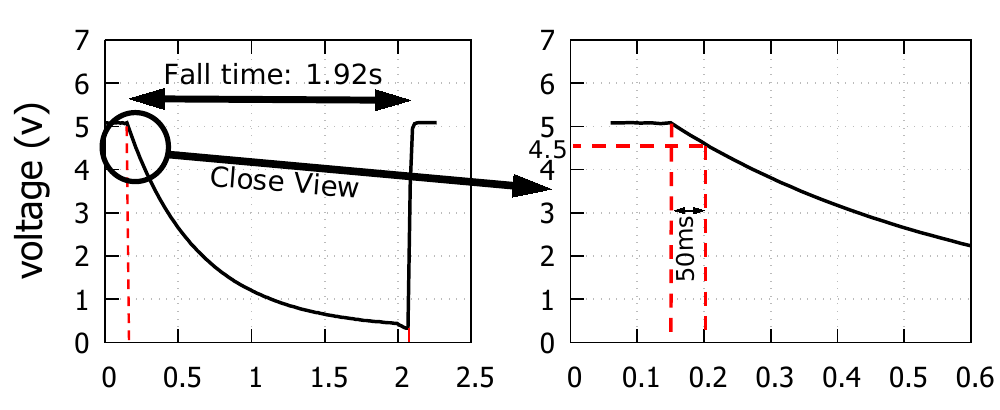}%
			\label{fig:withoutssd}}
		\hfil
		\hspace{-.8pt}
		\subfloat[When the PSU drives  two SSDs in RAID-1 configuration.]{\includegraphics[width=.45\textwidth]{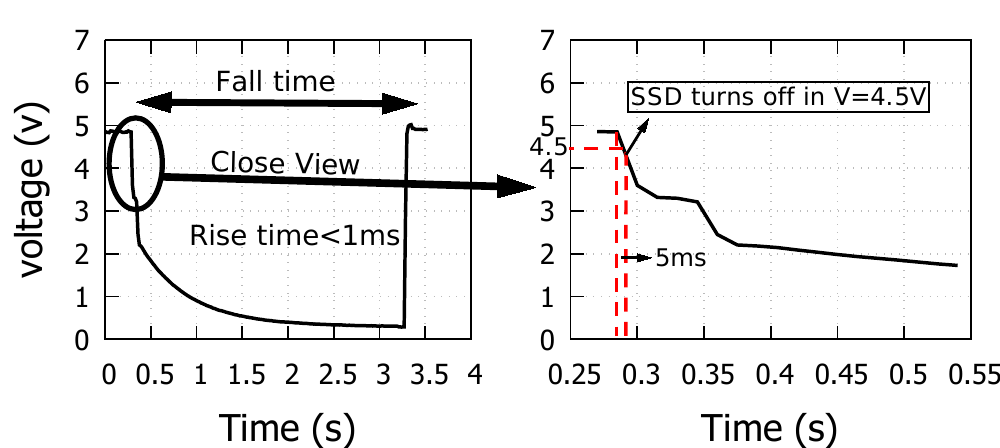}%
			\label{fig:withssd}}

		\caption{The output voltage of the PSU.}
		
		\label{fig:fall}
	\end{figure}

	The proposed reliability test platform injects the real physical fault. To inject real power outage fault, we model the real discharge delay of large size capacitors that are employed in the PSU. By conducting several {experiments,} we observed the impact of discharge delay of such capacitors on the input voltage of SSDs. As depicted in Fig. \ref{fig:fall}, we observed the output voltage of the PSU in different cases: 1) when the PSU drives no SSD (Fig. \ref{fig:withoutssd}) and 2) when the PSU drives two SSDs (Fig. \ref{fig:withssd}). The results of the experiments reveal that the full discharge (5v to 0) delay of the PSU takes about $1,900ms$ when it drives two SSDs. The SSDs {become} unavailable through the application layer (\SWM{} in \emph{Host System}) when the input voltage drops to 4.5v during $5ms$ after power fault injection.
	
	%
	%
	%
	
	\input{ssd-info}
	\section{Evaluations and Observations} \label{SEC:EXPR}
	In this {section,} we evaluate the reliability of I/O cache including SSDs in RAID-1 configuration and show the ratio of different types of failures (described in Section \ref{sec:failure_type}) under power outage and high temperature.
	To do so, we have conducted experiments using a realistic fault injection platform on more than ten enterprise SSDs from different vendors where the {detailed} technical information of the SSDs is provided in Table \ref{table:SSD-info}. Note that the SSDs from type C support \say{power loss data protection} and hence experience small number of failures compared to other SSD types.
	
	The detailed information about \emph{Host System} used as \SWM{} in the test platform is reported in Table \ref{table:host_info}.
	{To implement I/O cache, we use an open-source cache module, EnhanceIO, as a kernel module where the under test SSDs in RAID-1 configuration are used in the cache layer of HDD. 
	The I/O cache is configured with the \emph{Least Recently Used} (LRU) replacement policy and the block size is equal to 4KB. {The total cache size is equal to 100 GB (20\% of disk space) to create promotion and eviction operations on the cache. Note that in case of larger cache size, the cache has enough space to serve all accesses with minimum (near zero) miss ratio and eviction/promotion operation.}
	In the {following} experiments, the write policy is set to \emph{Write Back} (WB) {while} in Section \ref{policy_impact}, we configure the cache {policy} in three types of WB, \emph{Write Through} (WT), and \emph{Read Only} (RO) to compare the impact of cache write policy on the reliability.}
	We implement RAID-1 configuration using \emph{mdadm} as a software-based RAID management tool for Linux systems.
	
	\begin{table}[!h]
		\centering
		\caption{ HW and SW {specification} of \emph{Host System} used in reliability test platform.}
		\label{table:host_info}
		\begin{tabular}{|l|l|}
			\hline
			\multicolumn{2}{|l|}{\textbf{Hardware}}                          \\ \hline
			Motherboard & Z97-A from ASUSTeK  corp. \\ \hline
			CPU         & Intel(R) Core(TM) i5                      \\ \hline
			RAM         & 8GB DDR3 from Hynix Semiconductor         \\ \hline
			{\begin{tabular}[l]{@{}l@{}}HDD\\ (OS)\end{tabular}}
            & 7.2K RPM, 500GB from SEAGATE corp.         \\ \hline
			{\begin{tabular}[l]{@{}l@{}}HDD\\ (Disk Subsystem)\end{tabular}}         & IntelliPower, 500GB from Western Digital         \\ \hline
			Under test SSDs         & According to Table \ref{table:SSD-info}        \\ \hline
			\hline
			\multicolumn{2}{|l|}{\textbf{Software}}                          \\ \hline
			OS          & Ubuntu 17.04                              \\ \hline
			Kernel      & 4.10.0-19-generic                         \\ \hline
		\end{tabular}
	\end{table}
	
	We perform experiments and {examine} the impact of workload dependent parameters on the failure rate of the {SSD-based I/O cache (in RAID-1 configuration)} in presence of power outage {and high operating temperature}. We also study the impact of 1) workload WSS, 2) type of requests (read/write), 3) request size, 4) requested \emph{Input Output Operation per Second} (IOPS), 5) access pattern (random/sequential), and 6) sequence of the accesses (i.e., \emph{Read After Read} (RAR), \emph{Read After Write} (RAW), \emph{Write After Read} (WAR), and \emph{Write After Write} (WAR)).
	{In each experiment, we commit at least $24,000$ accesses and impose 600 faults into the SSDs. However, to investigate the impact of workload characteristics, we perform multiple experiments that increases the number of accesses and injected faults.}
	\footnote{{We mainly compare the number of failures per power fault that reveals no dependency between the number of fault injection and average number of data failures.}}
	In the following, {we elaborate} the experiments and report the impact of above-mentioned parameters on the failure rate and {the} ratio of different types of failures in presence of power outage and high temperature. 
	\subsection{Impact of Workload Working Set Size (WSS)}
	\label{sec:disksize}
	In this section, {we evaluate the reliability of I/O cache to \emph{examine} the impact of workloads WSS on the ratio of different types of failures.} To do so, we perform ten experiments by using the workloads with different {WSSs}. We change the workloads WSS from 1GB to 350GB and measure the failure ratio under power outage.{ In these experiments, the accesses are distributed in uniform random pattern and the requests size varies between 4KB and 1MB. We issue more than $96,000$ writes to the disk subsystem where the SSDs experience more than $2,400$ power failures during the experiments.}
	
\begin{figure}[h]
		\centering
		\includegraphics[scale=0.71]{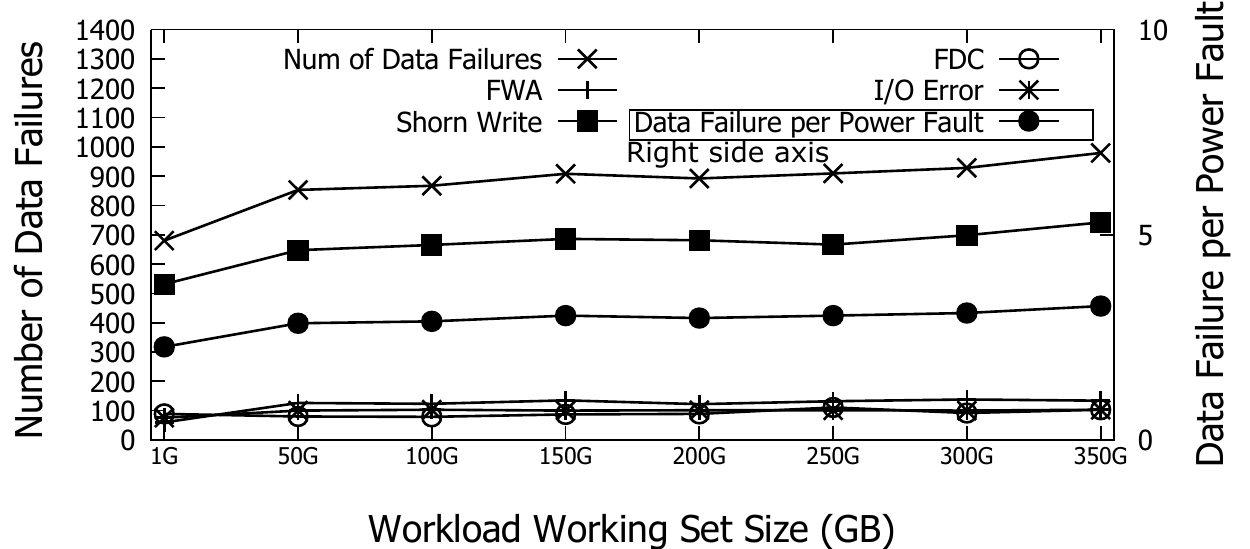}
		\vspace{-0.7em}
		\caption{{Impact of workload working set size on different types of failure.}}
		\vspace{-1em}
		\label{disk_size}
	\end{figure}

	The results of the experiments are shown in Fig. \ref{disk_size} (data failure per power fault is shown in the right side axis). We make two major observations: 1) {the failure rate in the I/O cache increases by 44\% when we increase the WSS from 1GB to 350GB under power failures that represents fewer failures in the workloads with smaller WSS.}
		 This is due to the fact that the time intervals between updating cache blocks in workloads with smaller WSS is less than the workloads with larger WSS. In this case, the written data resides in the cache for a short time where the probability of power outage in a such short time is low. 2) We do not experience unserilizable writes and flying writes failures during these experiments.
	
	We conclude that the workload with large WSS experiences more data failure rate compared to the workloads with smaller WSS. In this case, the failures from shorn write type become dominant.
	
	\subsection{Impact of Request Type}
	\label{sec:req_type}
	In this section, we conduct experiments to study the impact of request types of the workloads on the ratio of different types of failures {in the I/O cache}. To this end, we have performed five experiments where we change the percentage of read operations in the workloads  from 0 to 100\% and measure the failure ratio under power outage.{ In these experiments, the WSS of the running workloads is set to 380GB and the accesses are distributed in uniform random pattern. The requests size varies between 4KB and 1MB. We issue more than $150,000$ writes to the disk subsystem where the SSDs experience more than $3,000$ power failures during the experiments.}
	
\begin{figure}[h]
		\centering
		\includegraphics[scale=0.71]{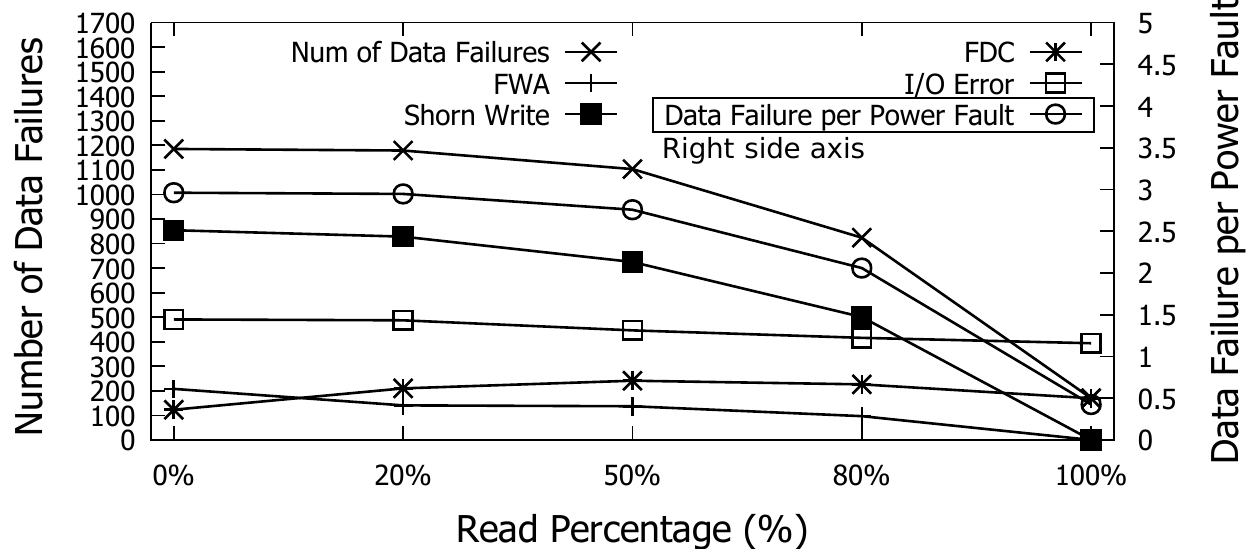}
		 \vspace{-0.7em}
		\caption{{Impact of request type on data failures.}}
		 \vspace{-1em}
		\label{read_percentage}
	\end{figure}

	The results of the experiments are shown in Fig. \ref{read_percentage} (data failure per power fault is shown in the right side axis). We make four major observations: {1) the failure rate decreases by increasing the number of read operations in the workloads.
		2) We observe failures only from FDC type in the 100\% read workload due to the failure of write operations during promoting missed data blocks into the I/O cache. 3) We observe no unserializable write and flying write types of failure.
		{The reason behind zero unserializable write failures is the workload type. We submit I/O requests with uniform random access pattern that the likelihood of submitting two consecutive write accesses in an identical address is low, and hence we do not observe any unserializable write failures.}
		 4) The shorn write failures are the dominant type in these experiments.}
	
	{We conclude that in the workloads with low number of read operations, the I/O cache experiences more data failures.
	In addition, in the workloads with large number of read operations, sudden power outage corrupts the data blocks of I/O cache (due to the corruption of write operations during promoting data blocks).}
	
	\subsection{Impact of Request Size}
	\label{sec:req_size}
	
	In this section, we evaluate the impact of request size of the workloads on the ratio of different types of failures. To this end, we have performed five experiments where the request size of the workloads is fixed to 4KB, 16KB, 64KB, 256KB, and 1MB in each experiment, {and} then we measure the failure ratio under power outage. { In these experiments, accesses are distributed in uniform random pattern and the requests size varies between 4KB and 1MB. We issue more than $225,000$ writes to the disk subsystem where the SSDs experience more than $4,500$ power failures during the experiments.}
	
\begin{figure}[h]
		\centering
		\includegraphics[scale=0.71]{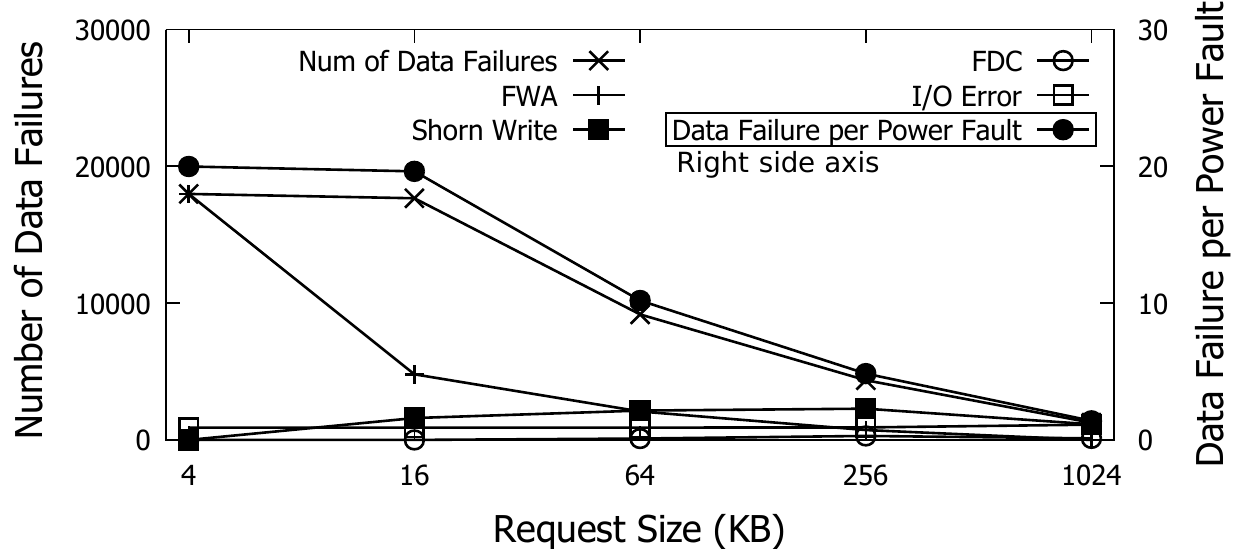}
		 \vspace{-0.7em}
		\caption{{Impact of request size on data failure.}}
		 \vspace{-1em}
		\label{req_size}
	\end{figure}

\begin{figure}[h]
		\centering
		\includegraphics[scale=0.71]{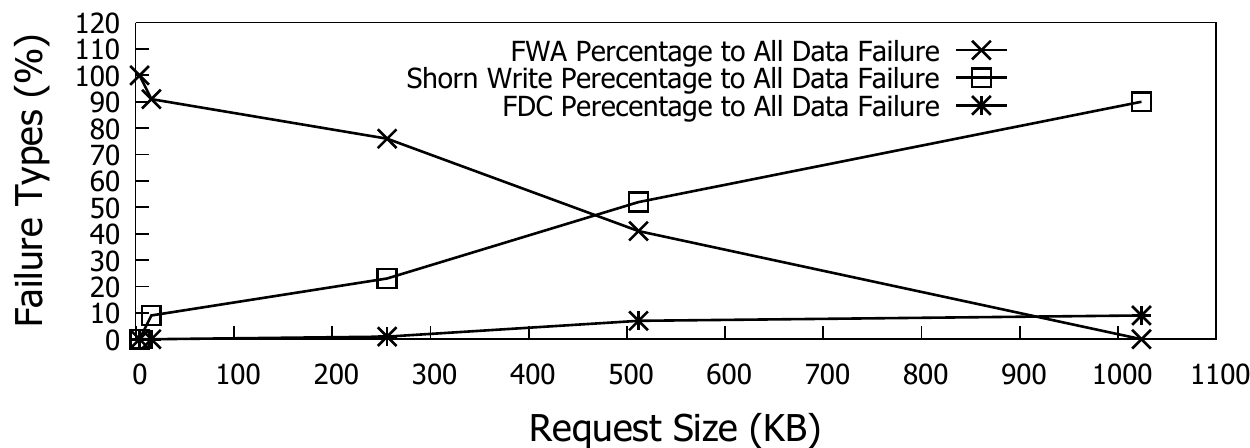}
		 \vspace{-0.7em}
		\caption{Impact of request size on data failure.}
		 \vspace{-1em}
		\label{Failure-perecentage}
	\end{figure}

	Fig. \ref{req_size} shows the results of these experiments (data failure per power fault is shown in the right side axis). We make three major observations: 1) the failure rate decreases by increasing the request size of the workloads. 2) In the workloads with smaller request size we do not observe shorn write while we experience large number of FWA failures ({we will elaborate this observation in the description of Fig.} \ref{Failure-perecentage}). 
	3) We observe {no failure from flying write and unserializable write types}.
	
	Fig. \ref{Failure-perecentage} shows the percentage of FWA, shorn write, and FDC failures based on the workloads request size. We observe that in the workloads with average request size less than {480KB,} the FWA failure is dominant while in the workloads with larger request size, the I/O cache experiences mainly shorn writes. This is because to the fact that in the workloads with small writes, large number of write pending request {is} buffered in {the SSD} {volatile elements within internal data path such as host FIFO buffer and DRAM buffer} \cite{cai2017error,cai2017errors,fifocomparison,meza2015large,luo2018architectural} and hence, in this case, power outage causes more FWA failures compared to the other type of failures. 
	{On the other hand, in case of submitting large size requests (larger than 500KB), the SSDs in the cache layer experience higher number of shorn writes than FWA. This observation also verifies the results of experiments presented in Fig. \ref{read_percentage}.} 
	{This is because shorn writes mainly occur due to an interrupt (here power outage) during write operation. For large size requests that the request is partitioned into multiple sub-requests, in case of power outage, the committed sub-requests before power outage will be completed while the operations during power outage are failed leading to shorn write failure.
		In contrast, small size requests finish and receive acknowledgment in a short time and the likelihood of power outage during small size write operation is much less than large size accesses. Thus, the number of shorn write failure in the workloads with large size requests is higher than the workloads with small size write accesses.}
	In addition, in the workloads with larger request {sizes,} we observe lower failures compared to workloads with smaller request size. {Finally, we observe increased failures from FDC type in the large size request.}
	
	We conclude that the workloads with small random write accesses experience large number of data failures under power outage compared to the workloads with large request sizes. 
	
	\subsection{Impact of Requests Access Pattern (Random/Sequential)}
	\label{sec:pattern}

{Here, we study the impact of workload access pattern on the ratio of different types of failures in the I/O cache configuration. To this end, we perform experiments with three different workloads including random and sequential access patterns (including write accesses with average request size equal to 512KB) and one real storage benchmark, Cello99 (14-Jan), mainly including 8KB partially sequential read and write accesses. Then we measure the failure ratio under power outage. We commit more than $26,000$ write requests to the disk subsystem and inject more than $200$ sudden power failures to the SSDs.}

\begin{figure}[h]
		\centering
		\includegraphics[scale=0.71]{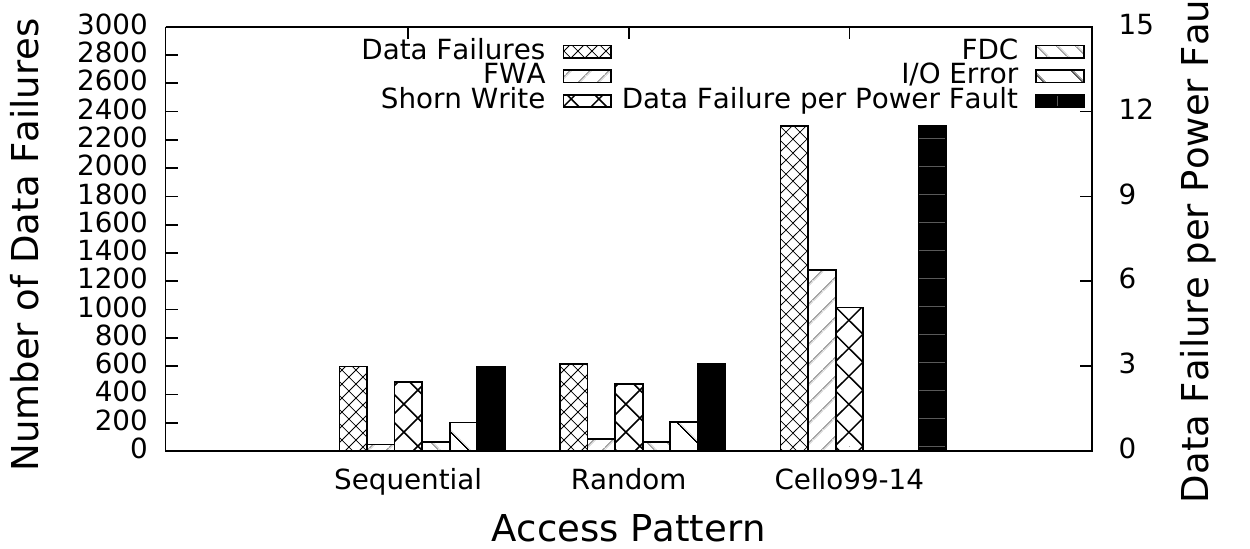}
		 \vspace{-0.7em}
		\caption{{Impact of access pattern on data failure (examining both synthetic and real storage workload)}}
		 \vspace{-1em}
		\label{access_pattern}
	\end{figure}

{Fig. \ref{access_pattern} shows the results of the experiments (data failure per power fault is shown in the right side axis). We make four major observations: 1) in the workloads with sequential access {pattern,} we experience {only} 2\% more data failure compared to the workloads with random accesses. 2) In all {workloads,} we observe larger number of FWA failures than shorn write failures. 3) We {observe} no unserializable write and flying write types of failure. 4) We observe higher range of failures (especially FWA) for Cello99 workload, which is mainly due to smaller requests size in this workload (8KB) compared to random and sequential workloads (512KB). This observation also verifies the results of experiments presented in Fig. 9 and Fig. 10.}

{We conclude that the access pattern of the workloads (sequential or random) has no significant impact on the failure rate of the I/O cache. In contrast, we observe a considerable impact of requests size on the failure ratio.}

%
%
%

	\subsection{Impact of IOPS}
	\label{sec:req_rate}
	In this section, we evaluate the impact of requested \emph{Input/Output Per Second} (IOPS) (i.e., the number of operations that are submitted {and responded by} the SSD in one second) {issued} by the workload on the ratio of different types of failures in the I/O cache (in RAID-1 configuration). To do so, we have performed five experiments where the requested IOPS of the workloads is fixed to 1.2K, 2.4K, 6K, 12K, and 20K in each experiment, then we measure the failure ratio under power outage. {The WSS of the experiments is set to 380GB and the requests size varies between 4KB and 1MB which are distributed in a uniform random pattern. In these experiments, we commit more than $120,000$ writes and inject more than $3,000$ power failures to the SSDs.} 
	
\begin{figure}[h]
		\centering
		\includegraphics[scale=0.71]{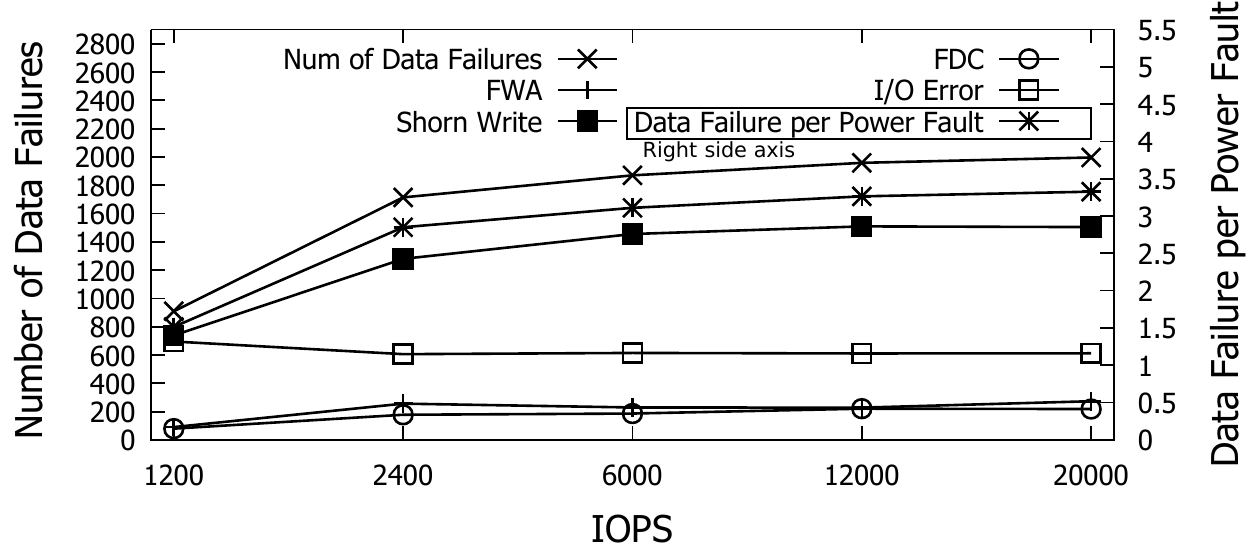}
		 \vspace{-0.7em}
		\caption{{Impact of requested IOPS submitted to the SSD on number of data failures.}}
		 \vspace{-1em}
		\label{req_rate}
	\end{figure}

	Fig. \ref{req_rate} shows the results of these experiments (data failure per power fault is shown in the right side axis). We make two main observations: {1) both responded IOPS from RAID configuration and failure rate are saturated when we increase the requested IOPS to 6K or greater.}
	2) The shorn write failure is dominant type of failure. 
	
	We conclude that in the workloads with high I/O load when the responded IOPS saturates the failure ratio saturates, respectively.

	\subsection{Impact of Sequence of the Accesses}
	\label{sec:req_seq}
	In this section, we conduct experiments to evaluate the impact of the workloads sequence of accesses on the reliability of I/O cache. To this end, we have performed experiments under power failures with different workloads where each workload mainly includes 1) RAW, 2) WAR, 3) RAR, and 4) WAW accesses. {The requests size is between 4KB and 1MB which the accesses are distributed in a uniform random pattern. In these experiments, $144,000$ write requests are committed to the subsystem where the SSDs experience $3,600$ sudden power outage.} 
	
\begin{figure}[h]
		\centering
		\includegraphics[scale=0.71]{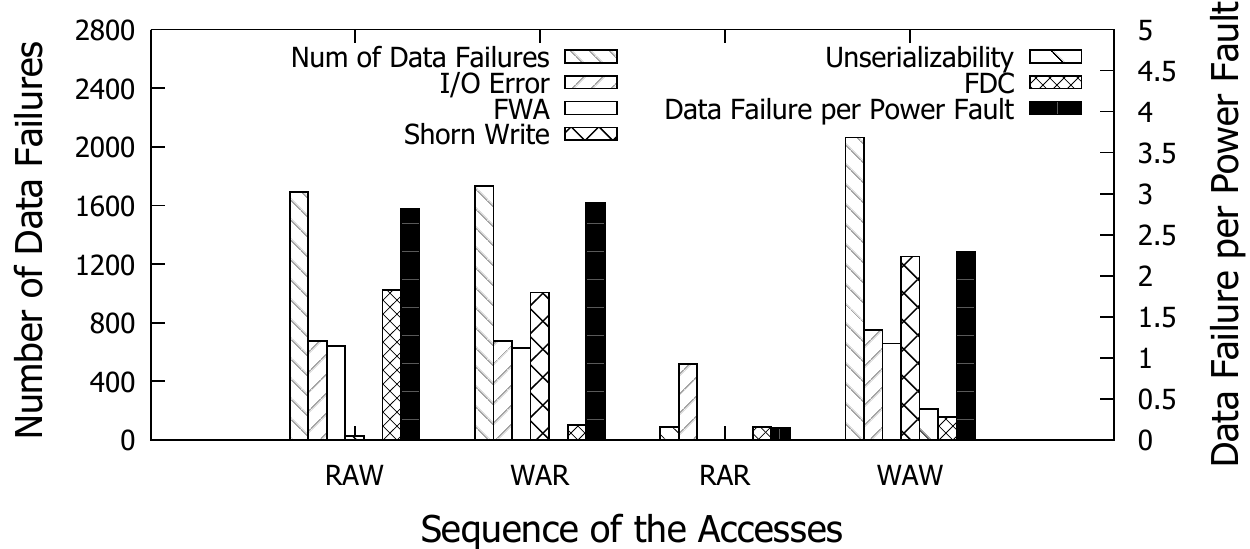}
		 \vspace{-0.7em}
		\caption{Impact of sequence of the accesses on data failure.}
		 \vspace{-1em}
		\label{sequence_of_accesses}
	\end{figure}

	Fig. \ref{sequence_of_accesses} shows the results of these experiments (data failure per power fault is shown in the right side axis). We make four major observations: 1) unserializable write failure occurs frequently in the workloads with WAW accesses. 2) We observe large number of shorn write failures in the workloads with WAR and WAW accesses. {3) The workloads with RAR accesses experience both I/O errors and data failures from FDC type under power outage.} 4)  FWA failure occurs in the workloads with RAW, WAR, and WAW accesses where the minimum and maximum number of such failure is observed in the workloads {with WAR and WAW accesses, respectively.}
	
	We conclude that the unserializable write failures mainly occur in the workloads with large number of WAW accesses under power outage.
	{In the workloads with RAR accesses, although there are no write requests from the application level, but we experience data failures due to the corruption of write operations during promoting data to the I/O cache.}
	{The shorn write failure occurs in all types of workloads while in the workloads with WAR and WAW accesses such failure becomes dominant.}
	
	\subsection{Impact of Disks Order in RAID-1 Configuration}
	\label{sec:disk_order}

	{In this section, we evaluate the impact of using SSDs from different vendors in different  orders in RAID-1, used in I/O cache configuration, on the ratio of failures.
		In a mirrored {(i.e., RAID-1)} configuration with two disks, the first disk is called \say{\emph{primary}} and  the second is called \say{\emph{secondary}} \cite{chen1994raid,thomasian2005performance,thomasian2006multilevel,thomasian2009higher,thomasian2006mirrored,thomasian2012performance,weddle2007paraid}. 
		To provide \emph{data consistency} in such subsystem, the controller periodically compares the written data in both disks. In case of any inconsistency between primary and secondary disks, the primary one is determined to store the valid data, and hence, the secondary disk will be updated with the primary disk's data. Furthermore, in RAID-1 configuration, write operations are {committed to} both primary and secondary disks, while read operations are \emph{only} supplied by {the} primary one. {A read} operation is provided by {the}  secondary disk when the primary one is busy due to supplying previous requests \cite{chen1994raid,thomasian2005performance,thomasian2006multilevel,thomasian2009higher,thomasian2006mirrored,thomasian2012performance,weddle2007paraid}.
		Considering two above-mentioned RAID-1 properties, using disks with different levels of reliability (i.e., from various vendors) as primary disk in mirrored configuration {can highly affect} the failure ratio. 
		To do so, we perform experiments under power failures with different RAID-1 configurations. In these {experiments,} we employ SSDs from various vendors providing different count of written logical blocks (i.e., different aging level) and different levels of reliability as {either} primary {or} secondary disks.}
	
\begin{figure}[h]
		\centering
		\includegraphics[scale=0.71]{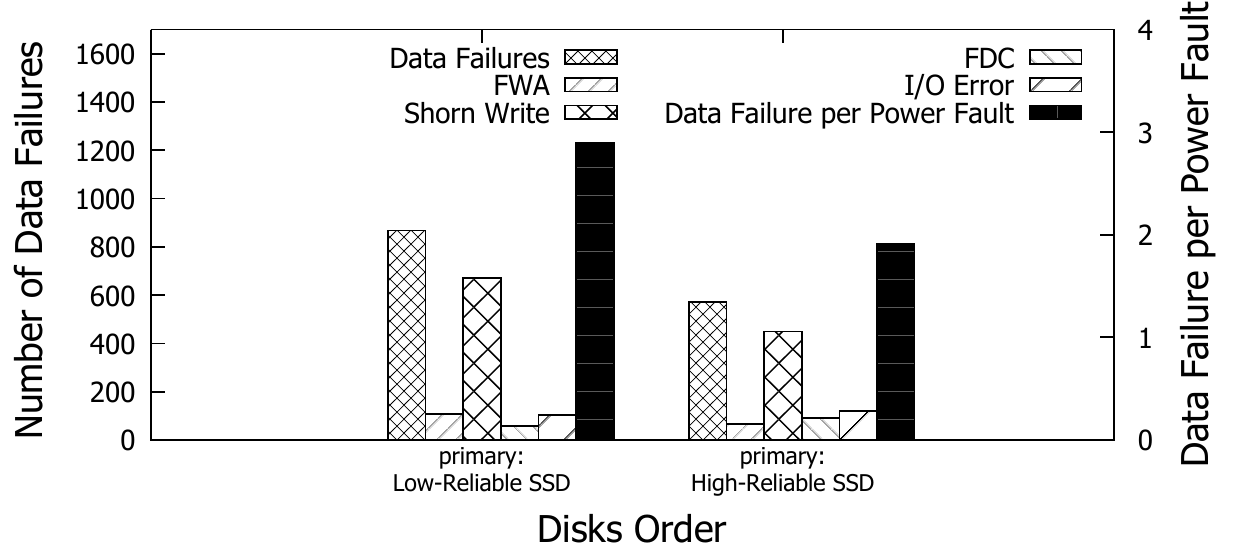}
		 \vspace{-0.7em}
		\caption{{Impact of disks order in RAID-1 Configuration.}}
		 \vspace{-1em}
		\label{disk_order}
	\end{figure}

	{In the experiments, we commit more than $48,000$ writes to the subsystem while the SSDs experience more than $1,200$ power failures.
		Fig. \ref{disk_order} shows the results of these experiments (number of data failures per power fault is shown in the right side axis). We observe that when we use a low reliable SSD (e.g., Type-A and Type-B) as  {the} primary disk, the failure ratio increases by 52\% (this is due to the fact that RAID-1 cannot mask total failures in the SSDs). In contrast, using SSDs from Type-D or Type-C as primary disk provides higher range of reliability by masking data failures occurred in the secondary disk that provides lower level of reliability. 
		We conclude that the ratio of data failure is {significantly} affected by the order of disks in I/O caches in RAID-1 configuration where employing a low reliable SSD as {the}  primary disk can increase the data failure about 52\%.}

	\subsection{Impact of High Temperature Faults}
	\label{sec:temp_result}
	In this section, we evaluate the reliability of {I/O cache} under high temperature failures.
	To this end, we perform experiments where the SSDs experience high temperature without any power failure. In the experiments, we commit more than $65,000$ random request to the SSDs. We increase the temperature of the SSDs up to 61 degrees Celsius (measured by SSDs S.M.A.R.T) and 64 degrees Celsius (measured by temperature sensor in the test platform). {The temperature of the SSDs in the experiments does not exceed the reported value in the datasheet.}
	
\begin{figure}[h]
		\centering
		\includegraphics[scale=0.71]{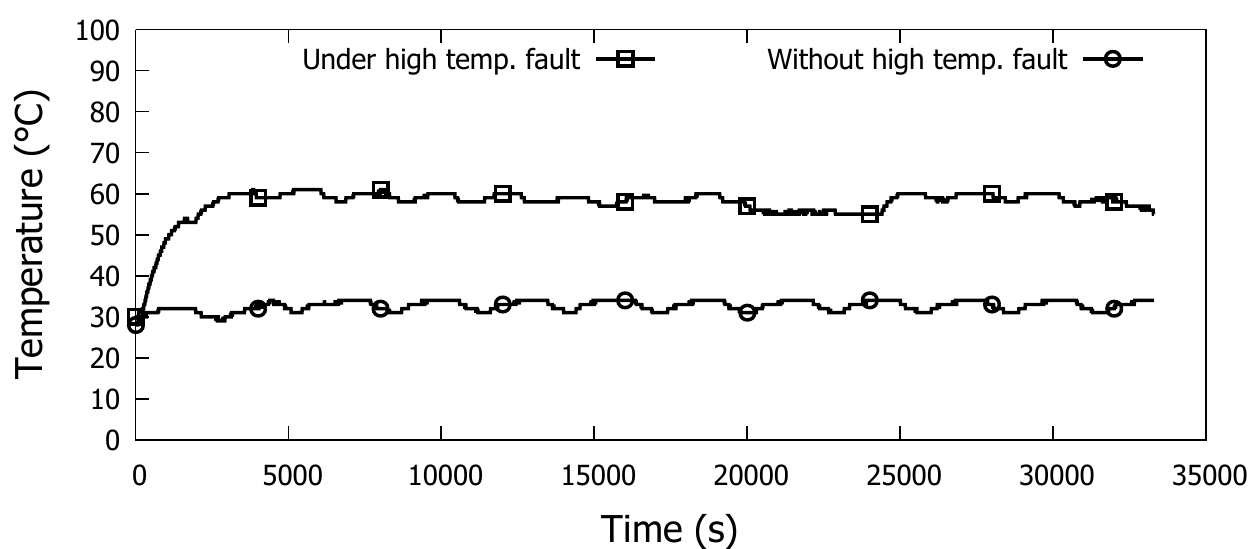}
		 \vspace{-0.7em}
		\caption{Disks temperature during experiments.}
		 \vspace{-1em}
		\label{disk_temp}
	\end{figure}

	Fig. \ref{disk_temp} shows the measured temperature of the SSDs during experiments. In this figure, we show the temperature of the SSDs in two cases: 1) during running the workload without any additional temperature faults and 2) during running the workload with injecting high temperature faults. It can be seen that, when SSDs supply the requests of the workload (described previously) the average temperature is about 31 degrees Celsius.
	
	{In the experiments,} we {observe} no data failure in the SSDs by increasing the temperature of the SSDs up to 64 degrees Celsius. {This is due to the fact that although high operating temperature increases the retention speed and failure rate, but existing enterprise SSDs reduce the number of accesses to the underlying SSDs (using throttling technique) resulting in reduced failure rate due to high temperature} \cite{meza2015large, luo2018architectural}.
	
	\subsection{Impact of I/O Cache Write Policy}
	\label{policy_impact}
	{In this section, we evaluate the reliability of I/O cache with different write policies under power outage. To this end, we conducted four experiments in which the cache write policy is set to 1) \emph{Write Back} (WB), 2) \emph{Write Through} (WT), and 3) \emph{Read Only} (RO). We examine the RO cache with two types of workloads including 100\% write (i.e., 0\% read) and 50\% read. The WSS of the workloads in all experiments is equal to 380GB. We commit more than $96,000$ requests to the disk subsystem where the size of accesses is between 4KB and 1MB with {a} uniform random access pattern. In each {experiment,} the SSDs experience more than $2,400$ sudden power failures.}
	
\begin{figure}[h]
		\centering
		\includegraphics[scale=0.71]{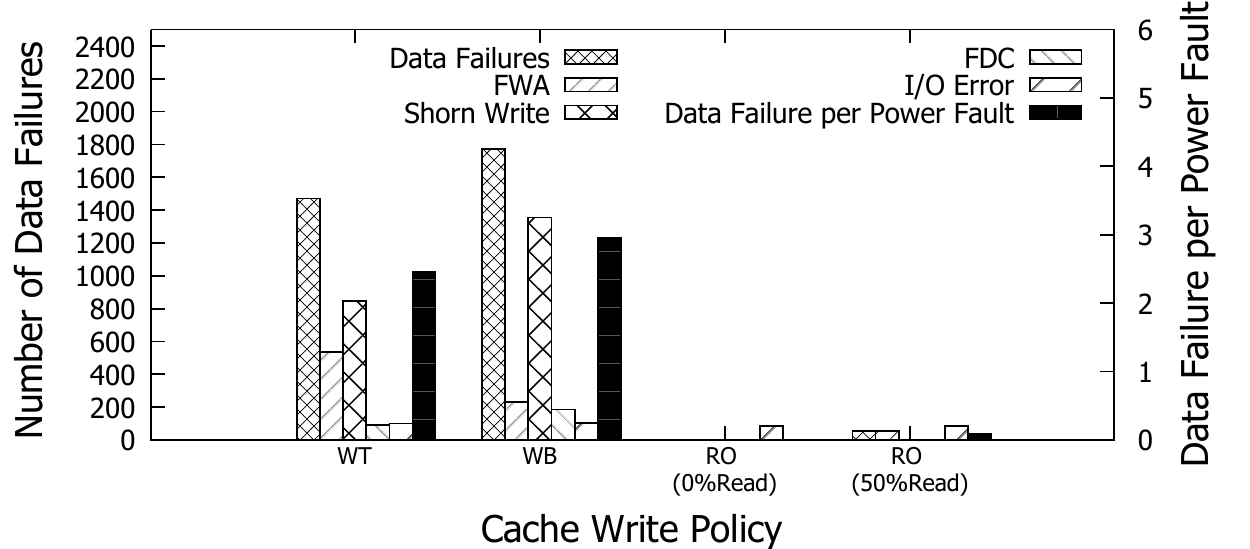}
		 \vspace{-0.7em}
		\caption{{Impact of cache write policy on the failure rate.}}
		 \vspace{-1em}
		\label{write_policy}
	\end{figure}

	{Fig. \ref{write_policy} shows the results of the experiments which reveals four major observations:
		1) failure rate in the WB cache is more than other cache policies {by 20\% (up to 32X)}. This is due to the fact that WB cache buffers all requests (read and write) where the write requests only reside in the cache until they become evicted. In a WB cache, if a data failure occurs on a data within the cache, such failed data will be evicted to the disk subsystem. Such condition affects the written data and also further read accesses which are responded from both cache and disk subsystem (i.e., hit or miss accesses).
		2) Although WT cache keeps two copies of data in both cache and disk, but data failure occurs {in such configuration} which cannot be recovered by the WT cache. In a WT cache, write requests first are supplied by disk subsystem and ACK is sent to the application, then the data is written to the cache for supplying future requests. In this case, if the written data in the cache {fails,} further read requests supplied from the cache {will} fail.
		3) WT cache experiences smaller failure rate compared to WB cache. The reason is that in the WT {cache,} there is no eviction from cache to the disk subsystem and hence, no corrupted data will be evicted to the disk. On the other hand, since WB cache keeps dirty blocks and evicts them to the disk subsystem, failed data in the cache {is} propagated to the disk subsystem.
		4) Although RO cache does not supply write requests, data failure occurs in the RO cache. In the experiment which we run the 0\% read workload on a RO cache, no read and write requests are supplied by cache (i.e., there is no accesses to the SSDs in the cache since all write requests are directed to the disk subsystem where there is no further read accesses to them). While in the second experiment with 50\% read workload, the RO cache directs writes to the disk subsystem while the read misses are buffered in the cache (i.e., are written in the cache). Sudden power outage during promoting the data to the cache may disrupt the write operation which leads to data failure. In this case, further read accesses which hit in the cache will fail.}
	
	{We conclude that the I/O cache in WB, WT, and RO policies experiences data failure where the failure rate in the WB cache is more than others {(by 20\%)}. Furthermore, if power failure occurs during promoting data to the cache, further read accesses will fail.}
	
	\subsection{Measuring the Current of SSDs}
	\label{sec:currency}
	
\begin{figure}[h]
		\centering
		\subfloat[Disks current during 10 hours of experiments (including 17 test intervals).]{\includegraphics[width=.47\textwidth]{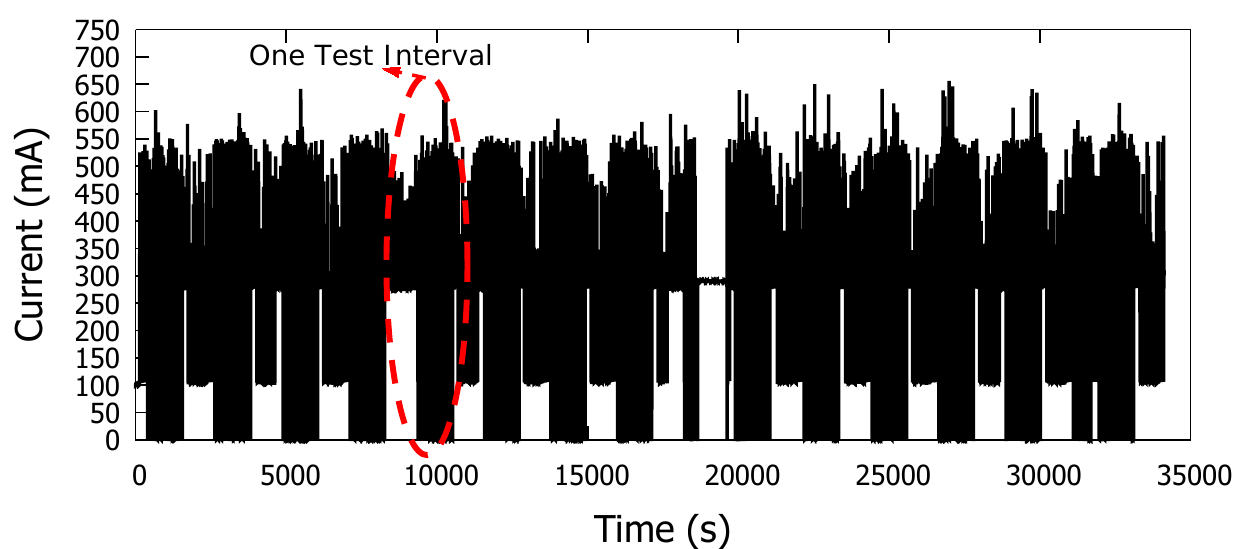}%
			\label{fig:current1}}
		\hfil
		\hspace{-.8pt}
		\subfloat[Disks current during one test interval (indicating read, write, and power outage phases).]{\includegraphics[width=.47\textwidth]{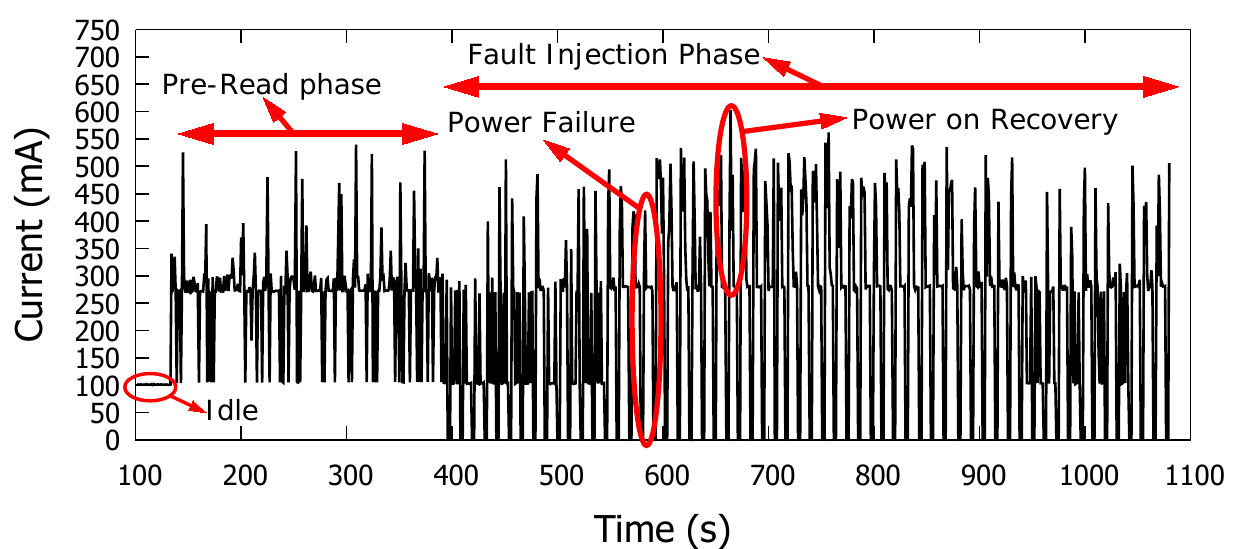}%
			\label{fig:current2}}
		
		\caption{Disks current during experiments.}
		
		\label{fig:disk_currency}
	\end{figure}

	{In this section,} the current of the SSDs is measured during 10 hours of experiments using INA219 module. In this {experiment,} we commit more than $60,000$ write requests to the {disk subsystem with the SSD-based I/O cache}.
	Fig. \ref{fig:disk_currency} shows the current of the SSDs in two cases: 1) during 10 hours (as depicted in Fig. \ref{fig:current1}) and 2) only {the}  first hour (as depicted in Fig. \ref{fig:current2}).
	{Using this report, we find the state of the disk and check the operations performing on the disk (idle, on, off, read operation, write operation, power on recovery, and fault injection instances). Then we decide when to inject the fault or when to start the failure detection algorithm.}
	As shown in Fig. \ref{fig:current2}, in the first 2-minutes, SSDs are idle and the current is less than 100mA. Then from $t=120s$ to {$t=400s$,} we have a reading phase {(before each I/O request)} where we read all written data in the addresses that we will rewrite to calculate the checksum of previously written data into the SSDs. At $t=400s$, when the reading phase finishes, we start the fault injection interval. In each power {outage,} the current of the SSDs is equal to zero. After each power outage, we observe a power on recovery interval (as an internal operation of the SSDs), in which we wait for the disk to become available through the application layer.

	\subsection{Failure Characterization of I/O Caches in Enterprise Storage Systems}
		\label{sec:freq_imp_failure}
\begin{figure}[!htb]
		\centering
		\includegraphics[scale=0.6]{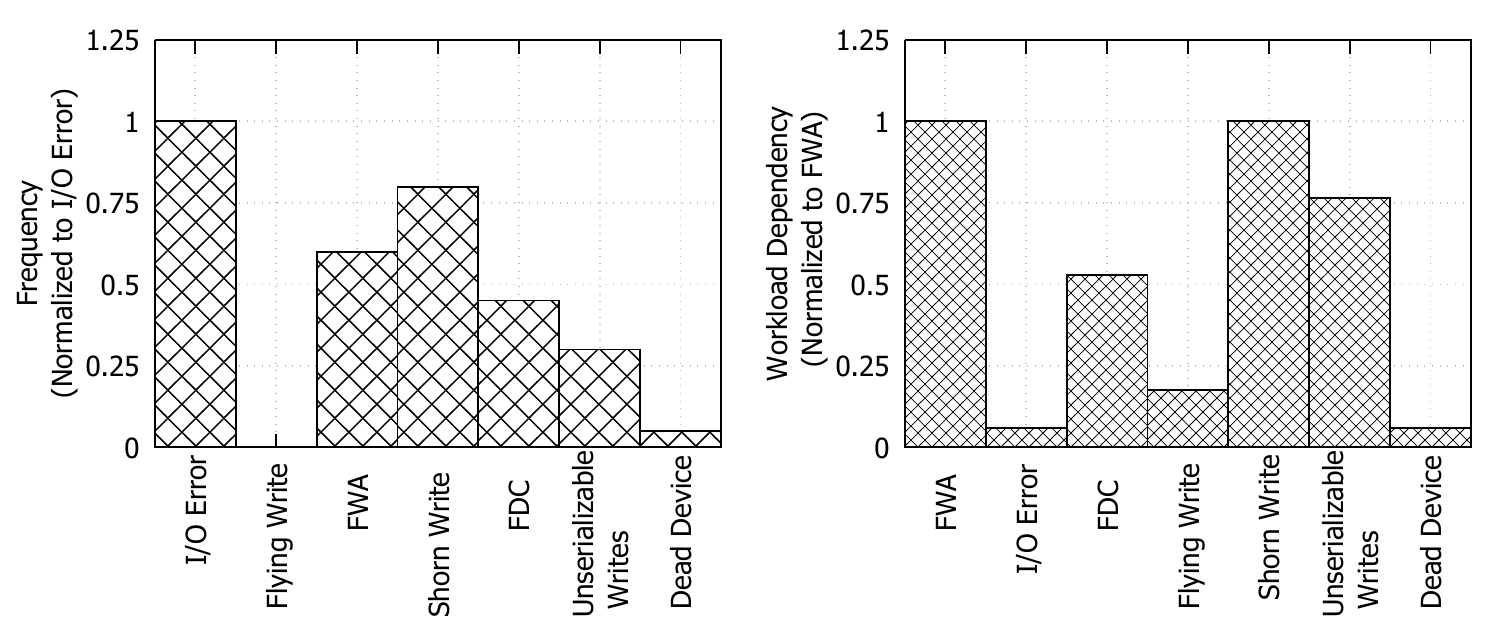}
		 \vspace{-0.7em}
		\caption{Frequency and importance of different types of failures.}
		 \vspace{-1em}
		\label{freq_imp_failure}
	\end{figure}

	In this section, we gather the results of previous experiments to indicate the frequency and {workload dependency}\footnote{{Workload dependency represents how different types of SSD failures are affected by different workload characteristics.}} of different types of failures (discussed in Fig. \ref{sec:failure_type}) as depicted in Fig. \ref{freq_imp_failure} and Table \ref{table:dependency1}. 
	{The main points of Fig. \ref{freq_imp_failure} and Table \ref{table:dependency1} are as follows:}
	\begin{enumerate}
		\item {As shown in Fig. \ref{disk_size}, we observe numerous {types} of SSD failures but \emph{none} of them are affected by varying workload WSS. In contrast, FWA, FDC, and shorn write failures are highly affected by request type.}
		
		\item {In the workloads with different {rates} of read accesses, the ratio of FWA varies more than 100\%. We observe similar behavior for FDC and shorn writes where the range of these failures varies respectively by 100\% and more than 100\% under different {ratios} of read accesses (Fig. \ref{read_percentage}).}
		
		\item {FWA and shorn writes are highly related to request size where they vary more than 100\% under different request sizes (Fig. \ref{req_size} and Fig. \ref{Failure-perecentage}), while request size has \emph{no} impact on other {types} of failures.}
		
		\item {FWA is partially affected by workload access pattern (less than 5\%). Similarly, we observe only a range of 5\% and 20\% variation for FDC and shorn write failures under different access patterns.}
		
		\item {FWA and FDC are highly related to requested IOPS of the running workload (by 100\%), while shorn write only varies by 75\% under different requested IOPS. We also observe other {types} of failures while they are not dependent to the requested IOPS of the workload.}
		
		\item {FWA, shorn write, and unserializable write failures are highly related to the sequence of the accesses. We observe significant (more than 100\%) variation in the ratio of unserializable writes under different {sequences} of accesses, where such failure only occurs in case of WAW accesses. Similarly, FDC is also highly related to sequence of accesses.}

		\item {We do \emph{not} observe any relation between I/O error and dead device failures with workload characteristics.}
		
		\item {In our experiments, we do \emph{not} observe flying write failure on the SSDs. However, we consider a small percentage of workload dependency for this type of failure compared to I/O error and dead device. This is because, flying writes mainly occur due to wrong destination address which seems to be related to workload characteristics.}
	\end{enumerate}
	

{According to {the} above-mentioned observations, we sort the dependency of different failures to the workload characteristics (Fig. \ref{freq_imp_failure}), where FWA and shorn writes are determined as the most workload dependent failures. Unserializable writes and FDC are respectively less workload dependent failures compared to two previous types. We determine the flying writes, I/O error, and dead device as the failures with \emph{zero} workload dependency.}

	\begin{table}[!htb]
		\caption{{Impact of workload characteristics on different types of SSD failures observed in our experiments.}}
		\label{table:dependency1}
		\tiny
		\centering
		\hspace*{-.25cm}
		\begin{tabular}{|l||c|c|c|c|c|c|c|}
			\hline
			\textbf{Section}                                    & \textbf{FWA}   & \begin{tabular}[c]{@{}c@{}}\textbf{I/O}\\\textbf{Error}\end{tabular} & \textbf{FDC}  & \begin{tabular}[c]{@{}c@{}}\textbf{Shorn}\\\textbf{Write}\end{tabular} & \begin{tabular}[c]{@{}c@{}}\textbf{Flying}\\\textbf{Write}\end{tabular} & \begin{tabular}[c]{@{}c@{}}\textbf{Unserial}\\\textbf{Write}\end{tabular} & \begin{tabular}[c]{@{}c@{}}\textbf{Dead}\\\textbf{Device}\end{tabular} \\ \hline
			Sec. \ref{sec:disksize}& -& - & - & - & - & - & - \\ \hline
			Sec. \ref{sec:req_type}& $> 100\%$  & - & 100\% & $> 100\%$  & - & - & - \\ \hline
			Sec. \ref{sec:req_size}& $> 100\%$ & - & -& $> 100\%$& -  & -& - \\ \hline
			Sec. \ref{sec:pattern}& $< 5\%$   & -  & $< 5\%$ & $< 20\%$ & -  & - & -  \\ \hline
			Sec. \ref{sec:req_rate}& 100\%  & - & $< 100\%$ & $> 75\%$ & -& - & -\\ \hline
			Sec. \ref{sec:req_seq}& $> 100\%$ & - & $> 100\%$ & $> 100\%$ & - & $>> 100\%$ & -\\ \hline\hline
			\begin{tabular}[c]{@{}l@{}}Workload\\Dep.\\ (1: Low,\\4: High)\end{tabular} & 4     & 1         & 2    & 4           & 1            & 3                    & 1           \\ \hline
		\end{tabular}
	\end{table}

	\section{Conclusion} \label{SEC:CONC}
	In this {paper,} we evaluated the reliability of  SSD-based I/O cache architectures in enterprise storage systems.
	To do so, we developed a Hardware-Software based reliability test platform for the SSDs that injects the physical failures such as power outage and high temperature faults that may occur commonly in large-size datacenters. The proposed test platform measures current, temperature, and power consumption of the SSDs and detects various types of failure that may occur in the SSDs and RAID configuration. We recognized different types of failures namely: \emph{False Write Acknowledge} (FWA), unserializable writes, {\emph{Full Data Corruption} (FDC)}, shorn writes, flying writes, I/O error, and dead device and measured the failure ratio in different conditions. We evaluated the impact of workload dependent parameters such as workload \emph{Working Set Size} (WSS), request size, request type, access pattern, requested I/O load, and sequence of the accesses on the reliability of SSD-based I/O caches in RAID-1 configuration. We conducted extensive experiments with various enterprise SSDs from top ten enterprise vendors and observed that the failure ratio in SSD-based {I/O cache architecture} under power outage is highly related to the I/O‌ parameters of the workload such as requests size and WSS of the requests while other parameters such as access patterns have no impact on the failure rate. We observed no data failure in the SSDs upon high temperature faults.  Furthermore, we observed that despite the high reliability of RAID-1 configuration, it fails in case of data inconsistency between disks which frequently happens upon power outage in data centers.

	\section*{Acknowledgement}
	This work has been partially supported by \emph{Iran National Science Foundation} (INSF) under grant number 9606071 and by HPDS Corp.

	{\small
		\bibliographystyle{IEEEtran}
		\bibliography{IEEEabrv,References}
	}
	
	\vskip -0.5\baselineskip plus -1fil
	\begin{IEEEbiography}[{\includegraphics[width=1in,height=1.25in,clip,keepaspectratio]{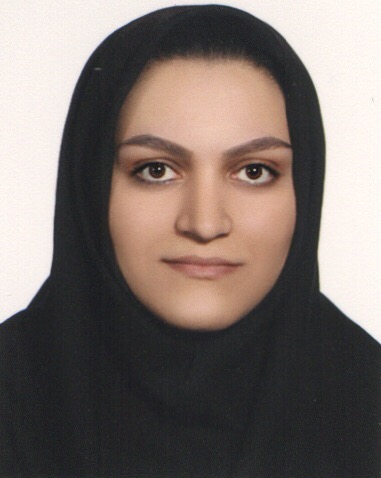}}]{Saba Ahmadian}
		received the B.Sc. and M.Sc. degrees in computer engineering from \emph{Sharif University of Technology} (SUT), Tehran, Iran, in 2013 and 2015, respectively. From 2011 to 2012, she was a member of \emph{Energy Aware Systems} (EASY) Lab, SUT, where she researched on power reduction techniques on embedded CPUs. From 2012 to 2015, she was a member of  \emph{Embedded Systems Research} (ESR) Lab, SUT, where she researched on low power and reliability-aware techniques on Automata-based embedded systems. Currently, she is a Ph.D. candidate at \emph{Data Storage, Networks, and Processing} (DSN) Lab at SUT under supervision of Dr. Hossein Asadi. Her research interests include storage systems design, virtualization platforms, fault tolerant design, and low power systems design.
		
	\end{IEEEbiography}
	\vskip -1\baselineskip plus -1fil
	\begin{IEEEbiography}[{\includegraphics[width=1in,height=1.25in,clip,keepaspectratio]{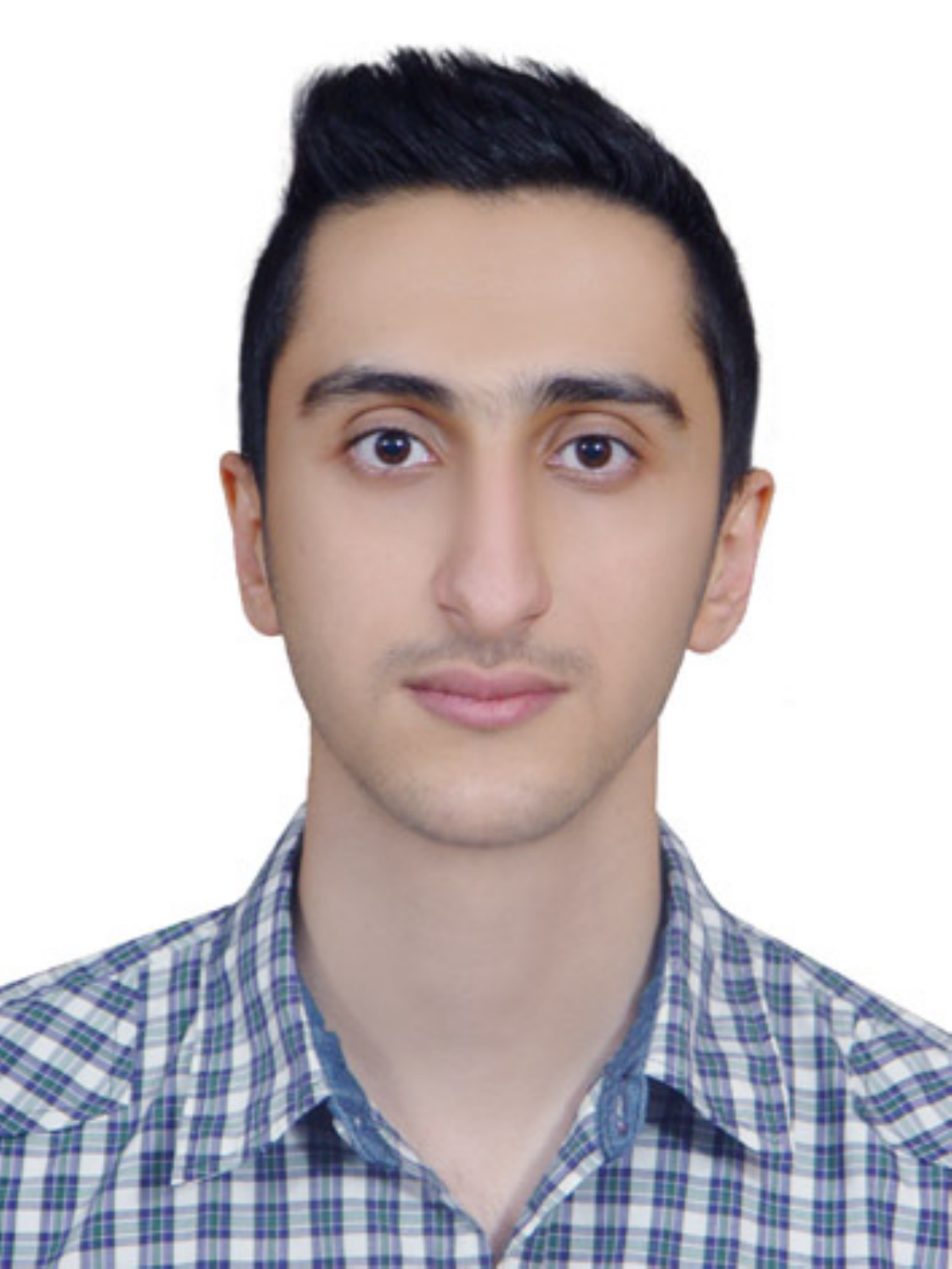}}]{Farhd Taheri}
		received the B.S. degree in computer engineering from Shahid Bahonar University, Kerman, Iran, in 2016 and M.Sc. degree at DSN Lab in SUT under supervision of Dr. Hossein Asadi.
		Currently, he is a Ph.D. student at \emph{Smart and Secure Systems} (3S) Laboratory at SUT under supervision of Dr. S. Bayat-Sarmadi.
		His research interests include Hardware security, Side Channel Attack, Solid-State Drives and fault tolerant design.
		
	\end{IEEEbiography}
	
	\vskip -1\baselineskip plus -1fil
	\begin{IEEEbiography}[{\includegraphics[width=1in,height=1.25in,clip,keepaspectratio]{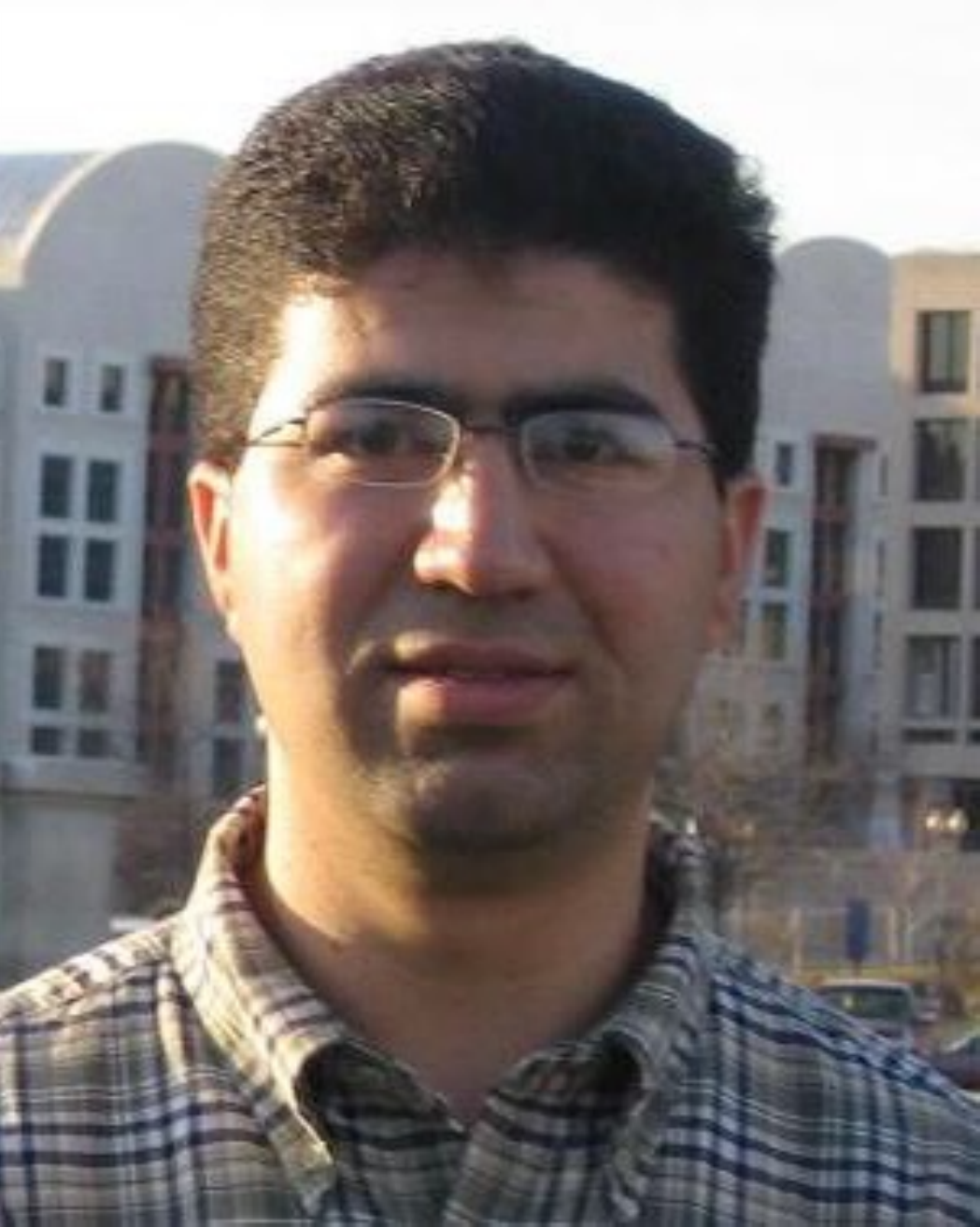}}]{Hossein Asadi}
		(M'08, SM'14) received the B.Sc. and M.Sc. degrees in computer engineering from the SUT, Tehran, Iran, in 2000 and 2002, respectively, and the Ph.D. degree in electrical and computer engineering from Northeastern University, Boston, MA, USA, in 2007. 
		
		He was with EMC Corporation, Hopkinton, MA, USA, as a Research Scientist and Senior Hardware Engineer, from 2006 to 2009. From 2002 to 2003, he was a member of the Dependable Systems Laboratory, SUT, where he researched hardware verification techniques. From 2001 to 2002, he was a member of the Sharif Rescue Robots Group. He has been with the Department of Computer Engineering, SUT, since 2009, where he is currently a tenured Associate Professor. He is the Founder and Director of the DSN Laboratory, Director of Sharif \emph{High-Performance Computing} (HPC) Center, the Director of Sharif \emph{Information and Communications Technology Center} (ICTC), and the President of Sharif ICT Innovation Center. He spent three months in the summer 2015 as a Visiting Professor at the School of Computer and Communication Sciences at the Ecole Poly-technique Federele de Lausanne (EPFL). He is also the co-founder of HPDS corp., designing and fabricating midrange and high-end data storage systems. He has authored and co-authored more than eighty technical papers in reputed journals and conference proceedings. His current research interests include data storage systems and networks, solid-state drives, operating system support for I/O and memory management, and reconfigurable and dependable computing.
		
		Dr. Asadi was a recipient of the Technical Award for the Best Robot Design from the International RoboCup Rescue Competition, organized by AAAI and RoboCup, a recipient of Best Paper Award at the 15th CSI International Symposium on \emph{Computer Architecture \& Digital Systems} (CADS), the Distinguished Lecturer Award from SUT in 2010, the Distinguished Researcher Award and the Distinguished Research Institute Award from SUT in 2016, and the Distinguished Technology Award from SUT in 2017. He is also recipient of Extraordinary Ability in Science visa from US Citizenship and Immigration Services in 2008. He has also served as the publication chair of several national and international conferences including CNDS2013, AISP2013, and CSSE2013 during the past four years. Most recently, he has served as a Guest Editor of IEEE Transactions on Computers, an Associate Editor of Microelectronics Reliability, a Program Co-Chair of CADS2015, and the Program Chair of CSI National Computer Conference (CSICC2017).
		
	\end{IEEEbiography}

\end{document}

%% file: ssd-info.tex
\begin{table*}[!htb]
	\centering
	\caption{Specification of employed SSDs in the experiments (DWPD: Driver Write Per Day, UBER: Unrecoverable Bit Error Rate, MTBF: Mean Time Between Failures, LDPC: Low Density Parity Check \cite{7887724}).}
	\label{table:SSD-info}
	\scriptsize
	\begin{tabular}{cccccccccccc}
		\hline
		\multicolumn{1}{|c|}{\begin{tabular}[c]{@{}c@{}}SSD\\ Type\end{tabular}} & \multicolumn{1}{c|}{Size}   & \multicolumn{1}{c|}{\$/GB} & \multicolumn{1}{c|}{\begin{tabular}[c]{@{}c@{}}Bit\\ per\\ Cell\end{tabular}} & \multicolumn{1}{c|}{\begin{tabular}[c]{@{}c@{}}Release\\ Year\end{tabular}} & \multicolumn{1}{c|}{\begin{tabular}[c]{@{}c@{}}Read/Write\\ IOPS (4KB)\end{tabular}} & \multicolumn{1}{c|}{\begin{tabular}[c]{@{}c@{}}Sequential \\ Read/Write\\ (MB/s)\end{tabular}} & \multicolumn{1}{c|}{DWPD} & \multicolumn{1}{c|}{\begin{tabular}[c]{@{}c@{}}Driver \\ Life Time\\ (TB)\end{tabular}} & \multicolumn{1}{c|}{UBER} & \multicolumn{1}{c|}{\begin{tabular}[c]{@{}c@{}}MTBF\\ (Million\\ Hours)\end{tabular}} & \multicolumn{1}{c|}{Other Features}                                                       \\ \hline \hline
		\multicolumn{1}{|c|}{A}        & \multicolumn{1}{c|}{120 GB}  & \multicolumn{1}{c|}{0.5}   & \multicolumn{1}{c|}{TLC}                                                      & \multicolumn{1}{c|}{2015}                                                   & \multicolumn{1}{c|}{60/70 K}                                                          & \multicolumn{1}{c|}{530/410}                                                                   & \multicolumn{1}{c|}{0.68} & \multicolumn{1}{c|}{90}               & \multicolumn{1}{c|}{-}    & \multicolumn{1}{c|}{1.5}                                                            & \multicolumn{1}{c|}{LDPC, SLC caching} \\ \hline
		\multicolumn{1}{|c|}{B}        & \multicolumn{1}{c|}{120 GB} & \multicolumn{1}{c|}{0.61}  & \multicolumn{1}{c|}{MLC}                                                      & \multicolumn{1}{c|}{NA}                                                     & \multicolumn{1}{c|}{11.5/52 K}                                                        & \multicolumn{1}{c|}{420/120}                                                                   & \multicolumn{1}{c|}{2.75} & \multicolumn{1}{c|}{354}              & \multicolumn{1}{c|}{-}    & \multicolumn{1}{c|}{1.0}                                                            & \multicolumn{1}{c|}{-}                                                             \\ \hline
		\multicolumn{1}{|c|}{C}        & \multicolumn{1}{c|}{2 TB}   & \multicolumn{1}{c|}{0.69}  & \multicolumn{1}{c|}{MLC}                                                      & \multicolumn{1}{c|}{2015}                                                   & \multicolumn{1}{c|}{95/28 K}                                                        & \multicolumn{1}{c|}{510/485}                                                                   & \multicolumn{1}{c|}{3.6}  & \multicolumn{1}{c|}{$12,320$}            & \multicolumn{1}{c|}{$10^{-17}$} & \multicolumn{1}{c|}{2.0}                                                            & \multicolumn{1}{c|}{Power loss data protection}                                    \\ \hline
		\multicolumn{1}{|c|}{D}        & \multicolumn{1}{c|}{2 TB}  & \multicolumn{1}{c|}{0.4}   & \multicolumn{1}{c|}{TLC}                                                      & \multicolumn{1}{c|}{2016}                                                   & \multicolumn{1}{c|}{93/24 K}                                                          & \multicolumn{1}{c|}{540/520}                                                                   & \multicolumn{1}{c|}{0.9}  & \multicolumn{1}{c|}{$3,200$}             & \multicolumn{1}{c|}{$10^{-17}$} & \multicolumn{1}{c|}{3.0}                                                            & \multicolumn{1}{c|}{-}                                                             \\ \hline
	\end{tabular}
\end{table*}

%% file: SSD-RAID-Reliability-2019-Sep-30.bbl
\begin{thebibliography}{10}
\providecommand{\url}[1]{#1}
\csname url@samestyle\endcsname
\providecommand{\newblock}{\relax}
\providecommand{\bibinfo}[2]{#2}
\providecommand{\BIBentrySTDinterwordspacing}{\spaceskip=0pt\relax}
\providecommand{\BIBentryALTinterwordstretchfactor}{4}
\providecommand{\BIBentryALTinterwordspacing}{\spaceskip=\fontdimen2\font plus
\BIBentryALTinterwordstretchfactor\fontdimen3\font minus
  \fontdimen4\font\relax}
\providecommand{\BIBforeignlanguage}[2]{{%
\expandafter\ifx\csname l@#1\endcsname\relax
\typeout{** WARNING: IEEEtran.bst: No hyphenation pattern has been}%
\typeout{** loaded for the language `#1'. Using the pattern for}%
\typeout{** the default language instead.}%
\else
\language=\csname l@#1\endcsname
\fi
#2}}
\providecommand{\BIBdecl}{\relax}
\BIBdecl

\bibitem{emc_fast_cache}
{Dell EMC Corp.}, ``{EMC UNITY: FAST TECHNOLOGY OVERVIEW},'' White paper,
  {Accessed: Jan. 2019}.

\bibitem{netapp_ssd_cache}
{NetApp}, ``{SSD Cache Feature},'' https://library.netapp.com, {Accessed: Jan.
  2019}.

\bibitem{hp_smart_cache}
{HP}, ``{HPE Smart Array SR SmartCache},'' https://h20195.www2.hpe.com,
  {Accessed: Jan. 2019}.

\bibitem{ahmadian2018eci}
S.~Ahmadian, O.~Mutlu, and H.~Asadi, ``{ECI-Cache: A High-Endurance and
  Cost-Efficient I/O Caching Scheme for Virtualized Platforms},''
  \emph{Proceedings of the ACM on Measurement and Analysis of Computing Systems
  (POMACS)}, vol.~2, no.~1, p.~9, 2018.

\bibitem{reca}
R.~Salkhordeh, S.~Ebrahimi, and H.~Asadi, ``{ReCA: An Efficient Reconfigurable
  Cache Architecture for Storage Systems with Online Workload
  Characterization},'' \emph{IEEE Transactions on Parallel \& Distributed
  Systems (TPDS)}, no.~7, pp. 1605--1620, 2018.

\bibitem{lbica}
S.~Ahmadian, R.~Salkhordeh, and H.~Asadi, ``{LBICA: A Load Balancer for I/O
  Cache Architectures},'' in \emph{to appear in Design, Automation Test in
  Europe Conference Exhibition (DATE)}, March 2019.

\bibitem{micheloni2012inside}
R.~Micheloni, A.~Marelli, and K.~Eshghi, \emph{{Inside Solid State Drives
  (SSDs)}}.\hskip 1em plus 0.5em minus 0.4em\relax {Springer Science \&
  Business Media}, 2012.

\bibitem{salkhordeh2015operating}
R.~Salkhordeh, H.~Asadi, and S.~Ebrahimi, ``{Operating System Level Data
  Tiering Using Online Workload Characterization},'' \emph{{The Journal of
  Supercomputing}}, vol.~71, no.~4, pp. 1534--1562, 2015.

\bibitem{li2017workload}
Y.~Li, B.~Shen, Y.~Pan, Y.~Xu, Z.~Li, and J.~C. Lui, ``{Workload-Aware Elastic
  Striping With Hot Data Identification for SSD RAID Arrays},'' \emph{IEEE
  Transactions on Computer-Aided Design of Integrated Circuits and Systems
  (TCAD)}, vol.~36, no.~5, pp. 815--828, 2017.

\bibitem{7887724}
L.~Zuolo, C.~Zambelli, A.~Marelli, R.~Micheloni, and P.~Olivo, ``{LDPC Soft
  Decoding with Improved Performance in 1X-2X MLC and TLC NAND Flash-Based
  Solid State Drives},'' \emph{IEEE Transactions on Emerging Topics in
  Computing (TETC)}, pp. 1--1, 2018.

\bibitem{leventhal2008flash}
{Leventhal, Adam}, ``{Flash Storage Memory},'' \emph{{Communications of the
  ACM}}, 2008.

\bibitem{tseng2011understanding}
H.-W. Tseng, L.~Grupp, and S.~Swanson, ``{Understanding the Impact of Power
  Loss on Flash Memory},'' in \emph{{Design Automation Conference
  (DAC)}}.\hskip 1em plus 0.5em minus 0.4em\relax ACM, 2011, pp. 35--40.

\bibitem{zheng2013understanding}
M.~Zheng, J.~Tucek, F.~Qin, and M.~Lillibridge, ``{Understanding the Robustness
  of SSDs under Power Fault.}'' in \emph{{File and Storage Technologies
  (FAST)}}, 2013, pp. 271--284.

\bibitem{ahmadian-ssd-rel-date}
S.~Ahmadian, F.~Taheri, M.~Lotfi, M.~Karimi, and H.~Asadi, ``{Investigating
  Power Outage Effects on Reliability of Solid-State Drives},'' in
  \emph{Design, Automation \& Test in Europe Conference \& Exhibition
  (DATE)}.\hskip 1em plus 0.5em minus 0.4em\relax IEEE, 2018, pp. 207--212.

\bibitem{chen1994raid}
{Chen, Peter M and Lee, Edward K and Gibson, Garth A and Katz, Randy H and
  Patterson, David A}, ``{RAID: High-Performance, Reliable Secondary
  Storage},'' \emph{{ACM Computing Surveys (CSUR)}}, 1994.

\bibitem{thomasian2005performance}
A.~Thomasian and C.~Liu, ``{Performance Comparison of Mirrored Disk Scheduling
  Methods with a Shared Non-Volatile Cache},'' \emph{Distributed and Parallel
  Databases}, vol.~18, no.~3, pp. 253--281, 2005.

\bibitem{thomasian2006multilevel}
A.~Thomasian, ``{Multilevel RAID Disk Arrays},'' in \emph{Proc. of the 23rd
  IEEE/14th NASA Goddard Conf. on Mass Storage Systems and Technologies}.\hskip
  1em plus 0.5em minus 0.4em\relax Citeseer, 2006.

\bibitem{thomasian2009higher}
A.~Thomasian and M.~Blaum, ``{Higher Reliability Redundant Disk Arrays:
  Organization, Operation, and Coding},'' \emph{ACM Transactions on Storage
  (TOS)}, vol.~5, no.~3, p.~7, 2009.

\bibitem{thomasian2006mirrored}
A.~Thomasian, ``{Mirrored Disk Rouing and Scheduling},'' \emph{Cluster
  Computing}, vol.~9, no.~4, pp. 475--484, 2006.

\bibitem{thomasian2012performance}
A.~Thomasian and Y.~Tang, ``{Performance, Reliability, and Performability of a
  Hybrid RAID Array and a Comparison with Traditional RAID1 Arrays},''
  \emph{Cluster Computing}, vol.~15, no.~3, pp. 239--253, 2012.

\bibitem{weddle2007paraid}
C.~Weddle, M.~Oldham, J.~Qian, A.-I.~A. Wang, P.~Reiher, and G.~Kuenning,
  ``{PARAID: A Gear-Shifting Power-Aware RAID},'' \emph{ACM Transactions on
  Storage (TOS)}, vol.~3, no.~3, p.~13, 2007.

\bibitem{cai2017vulnerabilities}
Y.~Cai, S.~Ghose, Y.~Luo, K.~Mai, O.~Mutlu, and E.~F. Haratsch,
  ``{Vulnerabilities in MLC NAND Flash Memory Programming: Experimental
  Analysis, Exploits, and Mitigation Techniques},'' in \emph{High Performance
  Computer Architecture (HPCA),}.\hskip 1em plus 0.5em minus 0.4em\relax IEEE,
  2017, pp. 49--60.

\bibitem{dell-cc}
{Dell}, ``{Dell PowerEdge RAID Controller Guide: Consistency Checks},''
  http://www.dell.com/support/manuals, {Accessed: Jan. 2019}.

\bibitem{7499876}
A.~Grossi, C.~Zambelli, P.~Olivo, P.~Pellati, M.~Ramponi, C.~Wenger,
  J.~Alvarez-Herault, and K.~Mackay, ``{An Automated Test Equipment for
  Characterization of Emerging MRAM and RRAM Arrays},'' \emph{IEEE Transactions
  on Emerging Topics in Computing (TETC)}, vol.~6, no.~2, pp. 269--277, 2018.

\bibitem{7494670}
P.~Pouyan, E.~Amat, and A.~Rubio, ``{Memristive Crossbar Memory Lifetime
  Evaluation and Reconfiguration Strategies},'' \emph{IEEE Transactions on
  Emerging Topics in Computing (TETC)}, vol.~6, no.~2, pp. 207--218, 2018.

\bibitem{meza2015large}
J.~Meza, Q.~Wu, S.~Kumar, and O.~Mutlu, ``{A Large-Scale Study of Flash Memory
  Failures in the Field},'' in \emph{ACM SIGMETRICS Performance Evaluation
  Review}, vol.~43, no.~1.\hskip 1em plus 0.5em minus 0.4em\relax ACM, 2015,
  pp. 177--190.

\bibitem{cai2015data}
Y.~Cai, Y.~Luo, E.~F. Haratsch, K.~Mai, and O.~Mutlu, ``{Data Retention in MLC
  NAND Flash Memory: Characterization},'' \emph{{High Performance Computer
  Architecture (HPCA)}}, pp. 551--563, 2015.

\bibitem{cai2017error}
Y.~Cai, S.~Ghose, E.~F. Haratsch, Y.~Luo, and O.~Mutlu, ``{Error
  Characterization, Mitigation, and Recovery in Flash Memory Based Solid-State
  Drives},'' \emph{{Proceedings of the IEEE}}, vol. 105, pp. 1666--1704, 2017.

\bibitem{cai2017errors}
------, ``{Errors in Flash-Memory-Based Solid-State Drives: Analysis,
  Mitigation, and Recovery},'' \emph{arXiv preprint arXiv:1711.11427}, 2017.

\bibitem{cai2012error}
Y.~Cai, E.~F. Haratsch, O.~Mutlu, and K.~Mai, ``{Error Patterns in MLC NAND
  Flash Memory: Measurement, Characterization, and Analysis},'' in
  \emph{{Proceedings of the Conference on Design, Automation and Test in Europe
  (DATE)}}, 2012, pp. 521--526.

\bibitem{cai2013error}
Y.~Cai, G.~Yalcin, O.~Mutlu, E.~F. Haratsch, A.~Crista, O.~S. Unsal, and
  K.~Mai, ``{Error Analysis and Retention-Aware Error Management for NAND Flash
  Memory.}'' \emph{{Intel Technology Journal}}, vol.~17, no.~1, 2013.

\bibitem{Boboila2010a}
S.~Boboila and P.~Desnoyers, ``{Write Endurance in Flash Drives: Measurements
  and Analysis},'' in \emph{{File and Storage Technologies (FAST)}}, 2010, pp.
  115--128.

\bibitem{narayanan2016ssd}
I.~Narayanan, D.~Wang, M.~Jeon, B.~Sharma, L.~Caulfield, A.~Sivasubramaniam,
  B.~Cutler, J.~Liu, B.~Khessib, and K.~Vaid, ``{SSD Failures in Datacenters:
  What? When? and Why?}'' in \emph{{Proceedings of the 9th ACM International on
  Systems and Storage Conference (SYSTOR)}}.\hskip 1em plus 0.5em minus
  0.4em\relax ACM, 2016, p.~7.

\bibitem{grupp2009characterizing}
L.~M. Grupp, A.~M. Caulfield, J.~Coburn, S.~Swanson, E.~Yaakobi, P.~H. Siegel,
  and J.~K. Wolf, ``{Characterizing Flash Memory: Anomalies, Observations, and
  Applications},'' in \emph{{Annual IEEE/ACM International Symposium on
  Microarchitecture (MICRO)}}.\hskip 1em plus 0.5em minus 0.4em\relax IEEE,
  2009, pp. 24--33.

\bibitem{schroeder2016flash}
B.~Schroeder, R.~Lagisetty, and A.~Merchant, ``{Flash Reliability in
  Production: The Expected and the Unexpected},'' in \emph{{File and Storage
  Technologies (FAST)}}, 2016.

\bibitem{chang2016disturbance}
Y.-M. Chang, Y.-H. Chang, T.-W. Kuo, Y.-C. Li, and H.-P. Li, ``{Disturbance
  Relaxation for 3D Flash Memory},'' \emph{IEEE Transactions on Computers},
  vol.~65, no.~5, pp. 1467--1483, 2016.

\bibitem{kim2007virtual}
S.-K. Kim, J.~Choi, D.~Lee, S.~H. Noh, and S.~L. Min, ``{Virtual Famework for
  Testing the Reliability of System Software on Embedded Systems},'' in
  \emph{{Proceedings of the 2007 ACM symposium on Applied computing
  (SAC)}}.\hskip 1em plus 0.5em minus 0.4em\relax ACM, 2007, pp. 1192--1196.

\bibitem{fifocomparison}
{D. Rollins, Micron Technology Inc.}, ``{A Comparison of Client and Enterprise
  SSD Data Path Protection},'' https://www.micron.com, {Accessed: Jan. 2019}.

\bibitem{luo2018architectural}
Y.~Luo, ``{Architectural Techniques for Improving NAND Flash Memory
  Reliability},'' Ph.D. dissertation, Seagate Technology, 2018.

\bibitem{ahci_intel}
{Intel Corp.}, ``{AHCI Specification for Serial ATA},''
  https://www.intel.com/content/www/us/en/io/serial-ata/ahci.html, {Accessed:
  Jan. 2019}.

\end{thebibliography}
